\documentclass{revtex4}
\usepackage{amssymb}
\usepackage{amsmath}
\usepackage{latexsym}
\def\setR{\mathbb{R}}
\def\setN{\mathbb{N}}
\def\setC{\mathbb{C}}

\def\1{\mathbf {Id} }
\newcommand{\sss}[1]{\scriptscriptstyle #1}

\def\y {{\cal Y}^{\nu}({\cal Z)}}

\makeatletter
\renewcommand\thesection{\@Roman\c@section}
\renewcommand\theequation{\@arabic\c@section.\@arabic\c@equation}
\makeatother

\begin{document}
\author{
T. Garidi$^{1}$} \affiliation{$1$ - LPTMC and F\'ed\'eration de recherche APC, Universit\'e Paris 7
Denis Diderot, boite 7020
F-75251 Paris Cedex 05, France.}
\email{garidi@kalymnos.unige.ch}
\title{What is mass in desitterian physics?}

\date{\today}

\begin{abstract}
In the present paper we discuss the relevance  for de Sitter
fields of the mass and spin interpretation of the parameters
appearing in the theory.   We show that these apparently
conceptual interrogations have important consequences concerning
the field theories. Among these,  it appeared that several authors
were using masses which they thought to be different, but which
corresponded to a common  unitary irreducible representation
(UIR), hence to identical physicals systems. This could actually
happen because  of the arbitrariness of their  mass definition in
the de Sitter (dS) space. The profound cause of confusion however
is to be found in the lack of connexion between the group
theoretical approach on the one hand, and the usual   field
equation (in local coordinates) approach on the other  hand. This
connexion will be established in the present paper and by doing so
we will get rid of any ambiguity by  giving a consistent and
univocal definition of a ``mass" term uniquely defined with
respect to a specific UIR of the de Sitter group.
\end{abstract}

\maketitle\newpage

\section{Introduction}
In the present paper we would like to give an unambiguous method
to identify a field in dS space. This appeared to be necessary
after  we noticed that various authors were describing fields
belonging to the same UIR although the involved masses were
different. This could only be due to different ways of introducing
the  mass parameters into the field equations . We feel that the
question of a preferred  mass definitions can only be answered in
reference to the flat case where mass and spin are well defined.
More precisely, we would like to show that the specificities of
the fields are best recognized according to their membership of
the carrier space of an UIR of the dS group. Indeed, contrary to
the mass values which after all (see section IV) seem to be
arbitrary, one finds that the concerned fields all belong to a
characteristic family of UIR's in the group representation
classification. Starting with a given UIR, we  then  will be able
to follow it in the flat limit (H$= 0$), owing to the group
representation contraction procedure  \cite{eric}. This provides
an efficient method to define a mass in dS space.

Before we actually more precisely point out why one should be
careful when working with  the mass and spin in dS space,  let us
first remind how one usually introduces these labels. Following a
minkowskian tradition, in almost every work one tends to
discriminate  the various dS fields according to their mass and
spin values. As stated in \cite{birrel} for scalar fields,
``..field quantization in curved spacetime proceeds in close
analogy to the minkowskian case. We start with the Lagrangian
density
$${\cal L}(x)=\frac{1}{2}\left(-g\right)^{1/2}\{
\left(g^{\mu\nu}\nabla_{\mu}\nabla_{\nu}\phi\right)-\left[m^2+\xi
R \right]\phi^2\}\,,$$ where $\phi$ is the scalar field and $m$
the mass of the field quanta. The coupling between  the scalar
field and the gravitational field represented by the term $\xi R
\phi^2$ where $\xi$ is a numerical factor and $R$ is the Ricci
scalar..". A generalized Lagrangian density is defined for fields
of arbitrary spin in curved space times  by working on the  flat
Lagrangian density ($\partial$ replaced by $\nabla$,..) but
without changing the mass parameter (see \cite{birrel}). This
however raises the question of knowing to which extent the two
imported entities (the spin  and $m^2$) are adapted to the dS
space.

Concerning the mass parameter, there are several facts discussed
in the following and which should make us at least suspicious
towards a minkowskian interpretation of $m$. First of all, it
appears that the mass parameter for higher spin fields features
remarkable properties first observed in \cite{deser1,deser2} in
the general framework of constant curvature spaces.  It has been
pointed out that for specific mass values a new gauge invariance
allows to reduce the degrees of freedom of the associated field
(partially massless field). These partially massless fields are
found at border values which separate unitary from non unitary
regions of the field. A consequence of the non unitary regions is
that  the mass range  admits holes corresponding to these non
 unitary representation of the
dS group. In Ref. \cite{higu2} this forbidden mass range
phenomenon has been discussed for spin-$2$ fields,  and more
recently \cite{deser3} the systematic appearance of these
forbidden  mass values   has been addressed in the case of higher
spin fields. In short, it turned out that very specific mass
values yielded very specific properties of the involved field
\cite{deser4,deser5,deser6}. We will show on several examples that
the mass values for these fields do not tell anything in
particular whereas they all belong to the same family of UIR.

Secondly, it is found that the use of a mass parameter  is
sometimes   misleading because it actually happens that negative
values possibly correspond to some dS unitary irreducible
representation \cite{novello1}. Contrary to the Poincar\'e group,
the dS group Casimir eigenvalues can take negative values. This
observation is the starting point of this paper. Indeed, the
common approach to dS field theory where a positive constant -the
mass- is introduced into the Lagrangian, fails to describe all the
UIR's of the dS group. In the same way as for instance a
minkowskian mass somehow restricted to strictly positive values
would fail to describe the massless case,  the introduced
parameter $m^{2}$ might set a lower bound (the ``massless" case,
$m^2=0$) even though a negative value of this same parameter still
corresponds to an UIR of the dS group. These negative values are
then simply ruled out or  one is forced to find an explanation for
a phenomenon which in our view is a simple consequence of a wrong
choice of mass parameter.  {\bf We therefore claim that a mass
parameter is improper if one cannot be ensured that it covers the
entire list of UIR's}.  We will illustrate this discussion with an
example concerning the mass of the graviton and the question of
its value  debated in \cite{novello1}.

Another puzzling aspect of the mass interpretation of $m$ in
constant curvature spaces  is that the fields associated to the
value $m=0$ is not trivially linked  to  conformal invariance or
light cone propagation \cite{deser1}. In fact, whereas gauge
invariance, helicities $\pm s$, light cone propagation or
masslessness  are essentially synonymous in flat space, the
situation is rather more complicated in (A)dS spaces. In
\cite{deser2} it is for instance shown that while gauge invariant,
the spin $\frac{3}{2},2$ fields do not propagate only on the null
cone for the AdS space. Moreover, as it is stressed in
\cite{deser2,dewitt},  ``A scalar field propagating according to
$\nabla_{\mu}\nabla^{\mu}\phi=0$ (no mass term ) scatters from the
background, thereby propagating both on and inside the local null
cones. In this sense, the field appears to be ``massive''.." Again
we see that  a mass interpretation  is really confusing.

On the other hand, and since at large radius the dS space is close
to the flat space, we want to be able to indicate when the
parameter $m$ can be viewed as a physical mass for the minkowskian
observer. It turns out that the appropriate  tool providing a
satisfying resolution of all the questions we have raised is the
systematic use of the group representation approach. Indeed, the
group representations contractions will enable us to compare the
dS representations parameters to those of the Poincaré group
whenever this makes sense. It is found \cite{gareta1} that not
every UIR of the dS group admits a Poincar\'{e} group
representation in the limit H$= 0$. Fortunately,  there are  UIR's
which contract toward the minkowskian massless fields. These UIR's
will be   unambiguously associated to  the dS massless fields and
will set a lower bound for our dS mass definition. Moreover as we
have already said, the mentioned partially massless fields seem to
all belong to a specific family of UIR's. Thus we are able to
characterize these new gauge fields and possibly predict their
occurrence.

Although the best would be not to use the term of mass for
desitterian fields, we nevertheless will introduce the entity
$m^{2}_{H}\geq 0$ (as it is usually done in most of the papers
concerned with desitterian physics) supposed to depend on the
curvature $H$ and which will be uniquely and explicitly determined
through the involved UIR. We will show that a reasonable mass
definition reads
\begin{equation}
m_{H}^{2}= \left[(p-q)(p+q-1)\right]H^2\quad\in\setR\,,
\end{equation}
where $2p\in \setN$ and $q\in\setC$ label the various dS  UIR's.
We will show later on that $m_{H}^{2}$ is real for every value of
$q$. This parameter will be considered as a mass only for those
UIR's which have a minkowskian interpretation in the $H=0$ limit
or which admit a unique extension to the conformal group
SO$(2,4)$. Note also that when   the limit $H=0$ can be defined,
the parameter $p$ remains constant whereas $q\sim i m/H$ where $m$
is the minkowskian mass. For these representations, the
representation parameters are connected to the familiar Poincar\'e
group representation parameters which are the mass and the spin.
Consequently, it is only for these representations that we will
speak of mass and spin for  the involved parameters. Our point of
view provides a nomenclature for desitterian fields, which are
divided into purely desitterian (no minkowskian interpretation)
and more familiar fields. This approach may seem a little
anthropomorphic but it is motivated since at large radius ($1/H$)
at least locally, the mesurable entities should match the
minkowskian ones.

Because our mass definition may be different from those given in
the literature,  we present a systematic procedure for identifying
dS fields starting with the field equation. Hence  we will be able
to compare the various involved parameters. Throughout this work,
the ``identity" of a field will be given by specifying the
associated unitary irreducible representation. This will be done
within  the framework of ambient space formalism, most adapted to
group theoretical matters because Casimir operators are simple to
express. We systematically will reduce the field equation to the
Casimir eigenvalue equation (with the Casimir operators $\langle
Q^{(1)}\rangle $) given by
\begin{equation}\label{eq:casimir}
\left(Q^{(1)}-\langle Q^{(1)}\rangle\right){\cal K}(x)=0\,.
\end{equation}

We thus  construct dS elementary systems (in Wigner's sense) in
perfect analogy with the minkowskian case. Recall that in the
minkowskian  case, the field equations are the Casimir equation
with eigenvalues $m^{2}$ and $s$ (the mass and the spin). However,
we insist that neither the mass nor the spin serve to classify the
dS UIR's, or to label the field. A further advantage of our dS
nomenclature is that its graphical representation in terms of the
values of $p$ and $q$ will allow to identify the various fields in
a straightforward way. Notably we will characterize the new fields
recently addressed in the series of paper due to S. Deser and A.
Waldron \cite{deser1,deser2,deser3,deser4,deser5,deser6} and
possibly predict their appearance for higher spins.

This paper is organized as follows. First we shall recall a
complete classification of the UIR's of the dS group in terms of
the eigenvalues of the Casimir operators (Section II). This
classification is the sum of the works found in \cite{dix,tak},
see related references therein. Since our goal is to connect our
approach to the field equations traditionally given in literature,
we must indicate how to relate the Casimir operators to the
covariant derivatives or other intrinsic entities (Section III)
and finally show how to consistently introduce a mass parameter
which includes every UIR (Section IV). We then give examples of
recent debates where our approach may contribute to clarify the
situation.

\setcounter{equation}{0}
\section{Classification of the unitary irreducible representations of the de Sitter group SO$_{0}(1,4)$.}

Let us first recall \cite{fr} that the de Sitter is conveniently
seen as a hyperboloid embedded in a five-dimensional Minkowski
space $M^{5}$
$$ X_H=\{x \in \setR^5;x^2=\eta_{\alpha\beta} x^\alpha     x^\beta
=-H^{-2}=-\frac{3}{\Lambda}\},$$ where $\eta_{\alpha\beta}=$
diag$(1,-1,-1,-1,-1)$, and with the minkowskian induced  metric
$ds^{2}=\eta_{\alpha\beta}dx^{\alpha}dx^{\beta}=g_{\mu\nu}^{dS}dX^{\mu}dX^{\nu},\;\;\mu=0,1,2,3$.
The $X^\mu$'s are  $4$ space-time intrinsic coordinates of the dS
hyperboloid and $\Lambda$ is the cosmological constant. A tensor
field ${\cal K}_{\eta_1..\eta_r}(x)$ on $X_H$ can be viewed as an
homogeneous function ${\cal K}(x)$ on $ M^5$ with an arbitrary
degree of homogeneity $\lambda$. In order to guarantee that ${\cal
K}(x)$ lies in the dS tangent space, one must require the
transversality condition
$$x\cdot {\cal K}(x)=0.$$
On the dS space we define the tangential (or transverse)
derivative $\bar\partial$ in the following way
\begin{equation}
\bar\partial_\alpha=\theta_{\alpha\beta}\partial^\beta=\partial_\alpha
+H^2x_\alpha x\cdot\partial,\quad\mbox{verifying}\quad
x\cdot\bar\partial=0.
\end{equation}
The tensor with components
$\theta_{\alpha\beta}=\eta_{\alpha\beta}+H^2x_{\alpha}x_{ \beta}$
is called the transverse projection   operator. It satisfies
$\theta_{\alpha\beta}\; x^{\alpha}=\theta_{\alpha\beta} \; x^{\beta}=0$.

The kinematical group of the de Sitter space is the $10$-parameter
group SO$_0(1,4)$ (connected component of the identity in
SO$(1,4)$ ), which is one of the two possible deformations of the
Poincar\'e group (the other one being SO$_0(2,3)$ ). The  unitary irreducible
representations (UIR) of SO$_{0}(1,4)$ are characterized by the
eigenvalues of the two Casimir operators $Q^{(1)}$ and $Q^{(2)}$.  This is because these operators commute with the action of the group generators and therefore are constant in each unitary irreducible representation. They read
\begin{equation}
Q^{(1)}=-\frac{1}{2}L_{\alpha\beta}L^{\alpha\beta},\qquad
Q^{(2)}=-W_{\alpha}W^{\alpha},\label{eq:cas}
\end{equation}
where
\begin{equation}
W_{\alpha}=-\frac{1}{8}\epsilon_
{\alpha\beta\gamma\delta\eta}L^{\beta\gamma}L^{\delta\eta},
\quad\mbox{with  10 infinitesimal generators}\quad
L_{\alpha\beta}=M_{\alpha\beta}+S_{\alpha\beta}.
\end{equation}
The orbital part $M_{\alpha\beta}$ reads
$$M_{\alpha\beta}=-i (x_\alpha\partial_\beta-x_\beta\partial_\alpha)
\qquad \mbox{with}\qquad Q_0^{(1)}\equiv -\frac{1}{2}
\;M_{\alpha\beta}M^{\alpha\beta}\,,$$
where the  operator $Q_0^{(1)}$ represents the pure scalar part in the action of
$Q^{(1)}$. In order to precise the action of the spinorial  part \cite{gaha} $S_{\alpha\beta}$ on  a field ${\cal K}$, one must treat separately   the integer and half-integer cases. Integer spin fields can be represented by  tensors of rank $r$ and the spinorial action reads
\begin{equation}
S_{\alpha\beta}{\cal
K}_{\eta_{1}..\eta_{r}}=-i\sum_{i}\left(\eta_{\alpha\eta_{i}}{\cal
K}_{\eta_{1}..({\eta_{i}}\leftrightarrow{\beta})..\eta_{r}}-\eta_{\beta\eta_{i}}
{\cal K}_{\lambda_{1}..({\eta_{i}}\leftrightarrow
\alpha)..\eta_{r}}\right). \label{eq:spi}
\end{equation}
Half-integer spin fields with spin $s=r+\frac{1}{2}$  are represented
by four component spinor-tensor  ${\cal K}_{ \eta_{1}..\eta_{r} }^{i}$ with $i=1,2,3,4$.
The spinorial action is then divided in two different parts
$$S_{\alpha\beta}^{(s)}=S_{\alpha\beta}+S_{\alpha\beta}^{(\frac{1}{2})}\qquad
\mbox{with}\qquad S_{\alpha\beta}^{(\frac{1}{2})}=-\frac{i}{4}\left[\gamma_{\alpha},\gamma_{\beta}\right]\;,$$
where the five matrices $\gamma_{\alpha}$ are determined by the relations
$$\gamma^{\alpha}\gamma^{\beta}+\gamma^{\beta}\gamma^{\alpha}=2\eta^{\alpha\beta}\qquad
\gamma^{\alpha\dagger}=\gamma^{0}\gamma^{\alpha}\gamma^{0}\;.$$

The operateur $S_{\alpha\beta}$ defined by (\ref{eq:spi}) acts upon the
tensors indexes $\eta_{1}..\eta_{r}$ and  $S_{\alpha\beta}^{(\frac{1}{2})}$
acts upon the spinor indexes  given by $i$.
The symbol $\epsilon_{\alpha\beta\gamma\delta\eta}$ holds for the
usual antisymmetrical tensor. In fact the UIR's may be labelled by
a pair of parameters $\Delta =(p,q)$ with $2p \in {\setN}$ and $q
\in {\setC}$, in terms of which the eigenvalues of $Q^{(1)}$ and
$Q^{(2)}$ are expressed as follows \cite{dix,tak}:
$$
{Q^{(1)}}=[-p(p+1)-(q+1)(q-2)]{\1}, \quad
{Q^{(2)}}=[-p(p+1)q(q-1)]{\1}.
$$
According to the possible values for $p$ and $q$, three series of
inequivalent unitary representations may be distinguished: the
principal, complementary and discrete series.

\vspace{0.3cm} \noindent {\bf The Principal series of
representations }\vspace{0.2cm}

Also called ``massive'' representations, they are  denoted  by
$U_{p,\nu}$, and labelled with \quad $\Delta=(p,q)=\left(p,{1\over
2}+i\nu\right)$ where
\begin{eqnarray}
&&p=0,1,2,\dots \quad {\rm and} \quad  \nu\geq 0 \quad {\rm
or},\nonumber\\
&&p={1\over 2},{3\over 2},\dots \quad \ \, {\rm and} \quad  \nu
>0.\nonumber
\end{eqnarray}
The operators $Q^{(1)}$ and $Q^{(2)}$ take respectively the
following forms:
$${Q^{(1)}} =\left[ \left( {9\over 4}+{\nu ^2} \right)-p(p+1) \right]\,
{\1},\qquad {Q^{(2)}} =\left[ \left( {1\over 4}+{\nu ^2}
\right)p(p+1) \right]\, {\1}.$$

\vspace{0.3cm}
\noindent
{\bf The  complementary series representations}\vspace{0.2cm}

The  complementary series is denoted by  $V_{p,\nu}$  with
$\Delta=(p,q)=(p,{1\over 2}+\nu)$ and
\begin{eqnarray}
&&p=0 \quad {\rm and} \quad  \nu\in \setR\ ,\ 0<|\nu|<{3\over 2}
\quad {\rm
or},\nonumber\\
&&p=1,2,3,\dots \quad {\rm and} \quad \nu\in{\setR}\ ,\
0<|\nu|<{1\over 2}.\nonumber
\end{eqnarray}
The operators
$Q^{(1)}$ and $Q^{(2)}$ assume the following values
$$
{Q^{(1)}} =\Bigl[ \bigl( {9\over 4}-{\nu ^2} \bigr)-p(p+1)
\Bigr]\,{\1}, \qquad{Q^{(2)}} =\Bigl[ \bigl( {1\over 4}-{\nu ^2}
\bigl)p(p+1) \Bigr]\, {\1}\,.$$

\vspace{0.3cm}
\noindent
{\bf The discrete series of representations }\vspace{0.2cm}

The elements of the  discrete series of representations are
denoted by $\Pi_{p,0}$ and $\Pi^{\pm}_{p,q}$ where the signs $\pm$ stand for the helicity.
The relevant values for the couple  $\Delta=(p,q)$ are
\begin{eqnarray}
&&p=1,2,3,\dots  \quad {\rm and} \quad q=p,p-1,\dots,1,0 \quad {\rm or},\nonumber\\
&&p={1\over 2},{3\over 2},\dots \quad {\rm and} \quad
q=p,p-1,\dots,\ {1\over 2}.\nonumber
\end{eqnarray}

Let us add a few precisions concerning the UIR's which extend to
the conformal group SO$_0(2,4)$. First recall that in our view,
these UIR's will  correspond to the massless fields in de Sitter
space. Masslessness will in fact be synonymous to conformal
invariance throughout this paper. In Ref. \cite{barutbohm}, the
reduction of the SO$_0(2,4)$ unitary irreducibles representations
to the de Sitter subgroup SO$_0(1,4)$ UIR's are examined. It is
found that the SO$_0(1,4)$ UIR's which can be extended to UIR's of
the conformal group are the following:
\begin{itemize}
\item [-] The scalar representation with $p=0$, $q=1$ and $\langle
Q^{(1)}\rangle=2$, which, in the above classification, belongs to
the {\bf complementary series} of UIR. In that case, the
SO$_0(2,4)$ representation remains irreducible when restricted to
the SO$_0(1,4)$ subgroup. \item [-] The UIR's characterized by
$p=q=\frac{1}{2},1,\frac{3}{2},2,..$, which correspond to some
terms of the {\bf discrete series} of UIR. For any values such
that $p=q$, there are two inequivalent unitary irreducible
representations of SO$_0(2,4)$ and both remain irreducible when
restricted to SO$_0(1,4)$. These two UIR's denoted
$\Pi^{\pm}_{p,p}$  differ in the sign of the parameter $k_0=\pm p$
connected to a subgroup SO$(3)$ and  there is no operator in
SO$_0(2,4)$ which changes the value of that sign.    Therefore
these two UIR's are distinguished by an entity which we are
allowed to name the helicity.

\end{itemize}

\begin{figure}[h]
\begin{center}

\begin{picture}(0,0)%
\includegraphics{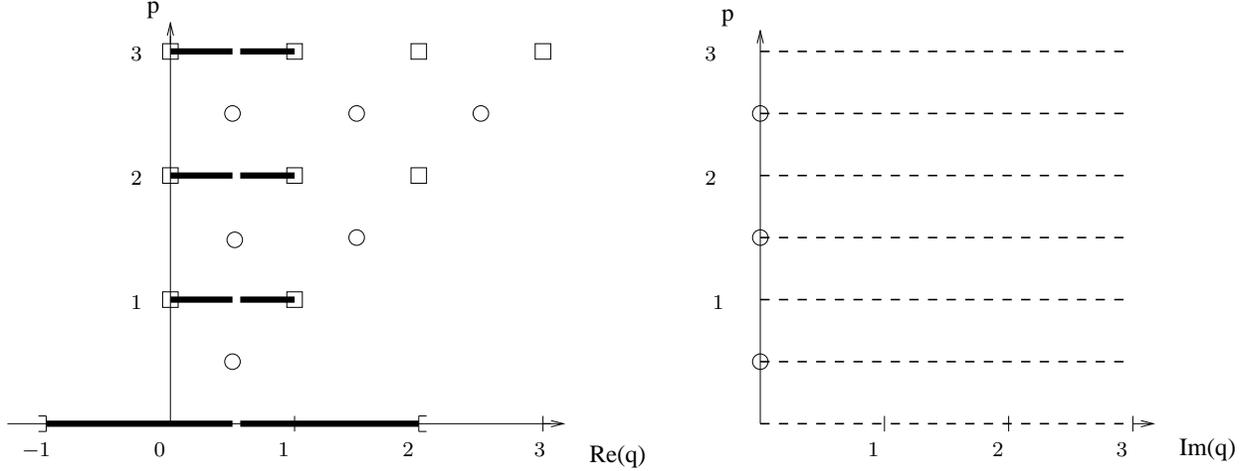}%
\end{picture}%
\setlength{\unitlength}{2565sp}%
\begingroup\makeatletter\ifx\SetFigFont\undefined
\def\x#1#2#3#4#5#6#7\relax{\def\x{#1#2#3#4#5#6}}%
\expandafter\x\fmtname xxxxxx\relax \def\y{splain}%
\ifx\x\y   
\gdef\SetFigFont#1#2#3{%
  \ifnum #1<17\tiny\else \ifnum #1<20\small\else
  \ifnum #1<24\normalsize\else \ifnum #1<29\large\else
  \ifnum #1<34\Large\else \ifnum #1<41\LARGE\else
     \huge\fi\fi\fi\fi\fi\fi
  \csname #3\endcsname}%
\else \gdef\SetFigFont#1#2#3{\begingroup
  \count@#1\relax \ifnum 25<\count@\count@25\fi
  \def\x{\endgroup\@setsize\SetFigFont{#2pt}}%
  \expandafter\x
    \csname \romannumeral\the\count@ pt\expandafter\endcsname
    \csname @\romannumeral\the\count@ pt\endcsname
  \csname #3\endcsname}%
\fi \fi\endgroup
\begin{picture}(11337,4561)(214,-4403)
\put(1426,-2836){\makebox(0,0)[lb]{\smash{\SetFigFont{8}{9.6}{rm}$1$}}}
\put(1426,-1636){\makebox(0,0)[lb]{\smash{\SetFigFont{8}{9.6}{rm}$2$}}}
\put(1426,-436){\makebox(0,0)[lb]{\smash{\SetFigFont{8}{9.6}{rm}$3$}}}
\put(8551,-4261){\makebox(0,0)[lb]{\smash{\SetFigFont{8}{9.6}{rm}$1$}}}
\put(9751,-4261){\makebox(0,0)[lb]{\smash{\SetFigFont{8}{9.6}{rm}$2$}}}
\put(10951,-4261){\makebox(0,0)[lb]{\smash{\SetFigFont{8}{9.6}{rm}$3$}}}
\put(7051,-2836){\makebox(0,0)[lb]{\smash{\SetFigFont{8}{9.6}{rm}$1$}}}
\put(6976,-1636){\makebox(0,0)[lb]{\smash{\SetFigFont{8}{9.6}{rm}$2$}}}
\put(6976,-436){\makebox(0,0)[lb]{\smash{\SetFigFont{8}{9.6}{rm}$3$}}}
\put(376,-4261){\makebox(0,0)[lb]{\smash{\SetFigFont{8}{9.6}{rm}$-1$}}}
\put(1651,-4261){\makebox(0,0)[lb]{\smash{\SetFigFont{8}{9.6}{rm}$0$}}}
\put(2851,-4261){\makebox(0,0)[lb]{\smash{\SetFigFont{8}{9.6}{rm}$1$}}}
\put(4051,-4261){\makebox(0,0)[lb]{\smash{\SetFigFont{8}{9.6}{rm}$2$}}}
\put(5326,-4261){\makebox(0,0)[lb]{\smash{\SetFigFont{8}{9.6}{rm}$3$}}}
\end{picture}

\caption{SO$_{0}(1,4)$ unitary irreducible representation
diagrams. Note that the right hand side diagram corresponds to the
Re$(q)=1/2$ plane.}\label{f1}
\end{center}
\end{figure}

We have pictured  these representations (up to $p=3$) in terms of
$p$ and $q$ on Figure \ref{f1} . The symbols $\bigcirc$ and
$\square$ stand for the discrete series with half-integer and
integer values of $p$ respectively. The complementary series is
represented in the same diagram  by bold lines. The principal
series is represented in the Re$(q)=1/2$ plane by dashed lines. We
have superposed  the three discrete series of representation with
values $p=1/2,3/2,5/2$ , Re$(q)=1/2$ and Im$(q)=0$ to the
principal series in order to show how these two diagrams fit
together. Note that the substitution $q\rightarrow (1-q)$ does not
alter the eigenvalues; the representations with labels
$\Delta=(p,q)$ and $\Delta=(p,1-q)$ are  said to be ``Weyl
equivalent". The weyl equivalent points can be localized on figure
6 starting from the points $q=\frac{1}{2}$ and $p=0,1,2,\dots$:
the bold lines (complementary series here) on the right hand side
of these points are weyl equivalent to the bold lines on the left
hand side, including the limiting points belonging to the discrete
series in the case $p>0$.

\setcounter{equation}{0}
\section{Minkowskian observer point of view}
First of all, let us recall that the Minkowski spacetime is obtained from the de Sitter spacetime in the $H=0$ limit. One easily shows that the dS metric tends toward the Mikowskian metric in that limit \cite{di}.
 A minkowskian  interpretation of the parameters $p,q$
is made possible when the group contraction method allows to
connect them to the Poincar\'{e} group parameters $s,m$. Recall
that the group contraction allows to follow a dS  UIR in the limit
$H=0$ \cite{eric,mini}. More precisely, one considers a family of
representations $U^{ H}$ of a group $G$ into  some spaces
${\cal H}_{ H}$ and a representation $U$ of a group $G'$ into a
space $\cal H$. The contraction procedure consists in giving a
precise meaning to the assertion $U^{H} \to U$ for $H\to 0$ (one
says that the representations $U^{ H}$ contract toward $U$).

For instance, a way of performing  the contraction of a dS UIR toward
a Poincar\'{e} group UIR denoted ${\cal P}({s,\pm m})$ (where $\pm$ stands for the positive and negative energies), will be to balance the vanishing $H$ with
the parameter $q$. Thus, a contraction by this method will only be
possible if $q$ has no upper bound which in view of our
classification corresponds to the principal series of UIR. Let us
now indicate more generally for which unitary irreducible
representation of the dS group such a procedure can be
implemented.

\vspace{0.2cm} \noindent {\bf \underline{de Sitter massive
fields}: }\vspace{0.2cm}

\begin{itemize}
\item
They correspond to  {\bf the principal series} of unitary
representation (also called the massive representations of the dS
group). In this case, the contraction procedure can be done by
having $p$ fixed, and the hole family of UIR's with $0<q<+\infty$
contracting toward the massive Poincar\'{e} group UIR. This is
achieved  with $q$ such that in the limit $H=0$, one gets $q\sim i
m/H$ \cite{eric,mini}. In the limit $H=0$, one gets
\begin{equation}
U_{ s,\frac{1}{2}+\frac{im}{H}}\to {\cal P}(s,\pm m)
\end{equation}
In the massive case,  the contraction is
performed with respect to the subgroup SO$_0(1,3)$ which is
identified as the Lorentz subgroup in both relativities, and the
concerned de Sitter  representations form the principal series.
They are precisely those ones which are induced by the {\it
minimal parabolic} \cite{lipsman} subgroup SO$(3) \times $SO
$(1,1) \times$(a certain nilpotent subgroup), where SO$(3)$
\underline{is} the space rotation subgroup of the Lorentz subgroup
in both cases. This fully clarifies the concept of spin in de
Sitter since it is issued from the {\it same} SO$(3)$. Thus the
principal series UIR's   contract toward the massive spin $s$
representations of the Poincar\'e group. The relevant tensor
field equation is given by
\begin{equation}\label{eq:principale}
\left[Q^{(1)}_{}- \left( {9\over 4}+{\nu ^2} \right)+s(s+1)
\right]{\cal K}=0\,.
\end{equation}
\end{itemize}

\vspace{0.2cm} \noindent {\bf \underline{de Sitter massless
fields}: }\vspace{0.2cm}

In that case we select those representations having a natural
extension to the conformal group SO$_0(2,4)$ and which are
equivalent to the massless spin $s$ UIR of the conformal extension  of the
Poincar\'{e} group \cite{anflafrons,barutbohm}.
\begin{itemize}
\item
For the scalar case, this UIR is found in {\bf the complementary
series} of unitary representation with the values $\Delta=(0,1)$.
It is also called conformally coupled massless case since it
corresponds to the conformally invariant field equation
\begin{equation}\label{eq:conforme}
\left(Q^{(1)}_{}- 2 \right)\phi=0\,,\qquad\Longrightarrow\qquad
\left(\Box_{H}+2H^2\right)\phi=0\,.
\end{equation}
In the limit $H=0$ they correspond to the massless scalar Poincar\'e group UIR
\begin{equation}
V_{0,1}^{}\to {\cal P}^{}(0,0)\,.
\end{equation}

\item
When $p\neq 0$, the only physical representations in the sense of
Poincar\'e limit are those with $p=q=s> 0$ which lie at the lower
end of {\bf the discrete series}. They are called the massless
representations of the dS group. In the limit $H=0$ they correspond to the massless spinorial Poincar\'e group UIR
\begin{equation}
\Pi_{s,s}^{\pm}\to {\cal P}^{\pm}(s,0)\,.
\end{equation}
The corresponding $r$-rank tensor
field equations read
\begin{equation}\label{eq:discrete2}
\left[Q^{(1)}_{}+2(s^2-1)\right]{\cal K} =0 \,.
\end{equation}
\end{itemize}

To summarize, we could say that in our point of view, the dS
massive fields with arbitrary spin $s$ correspond to the principal
series of unitary representations, and that the massless fields
correspond to the discrete series of unitary rep. with $p=q$
except for the scalar case which belongs to the complementary
series  with the values $p=0,q=1$. These are the fields for which
it is justified to use the terms of mass and spin. On the
contrary,  an example of a purely desitterian field is given by
the so-called ``massless" minimally coupled scalar field. This
field corresponds to the lowest term in the discrete series of
unitary representations (with $p=1,q=0$) and obeys the scalar
field equation $Q^{(1)}_0\phi=0$. This representation admits no
minkowskian interpretation.

\setcounter{equation}{0}
\section{Intrinsic field equations and unitary irreducible representations}

We now would like to adapt the above classification to the more
familiar language of the wave equations in local coordinates on
the manifold $X_{H}$ (intrinsic coordinates). This is actually a
simple way to identify every field given in literature on dS
space.  As we have said, the procedure consists in  reducing the
intrinsic wave equation to the eigenvalue equation
\begin{equation}\label{eq:casimir2}
\left(Q^{(1)}_{}-\langle Q^{(1)}_{}\rangle \right){\cal K}(x)=0\;,
\end{equation}
where  for integer spin fields, ${\cal K}(x)$ is a transverse ,
symmetric tensor of rank $r=s$. For half integer spin fields the
carrier space is made of transverse tensor spinor fields of rank
$r=s-\frac{1}{2}$.

The first step will be to rewrite   the intrinsic wave equations
in terms of ambient space notations. These  will help to make
explicit the group theoretical content, since they allow to easily
write the equation  in terms of the operator $Q^{(1)}_{}$.

The link between the ambient and the intrinsic notations is
provided by
\begin{equation}\label{passage1}
 h_{\mu_{1}..\mu_{r}}(X)=\frac{\partial
x^{\alpha_{1}}}{\partial X^{\mu_{1}}} ...\frac{\partial
x^{\alpha_{r}}} {\partial X^{\mu_{r}}}{\cal
K}_{\alpha_{1}..\alpha_{r}}(x)\,,
\end{equation}
and the covariant derivatives acting on a $r$-rank tensor are
transformed according to
\begin{equation}\label{passage2}
\nabla_{\mu}\nabla_{\nu}..\nabla_{\rho}h_{\lambda_{1}..\lambda_{r}}=
\frac{\partial x^\alpha}{\partial X^\mu} \frac{\partial x^\beta}
{\partial X^\nu}...\frac{\partial x^\gamma}{\partial X^\rho}
\frac{\partial x^{\eta_{1}}}{\partial X^{\lambda_{1}}}
...\frac{\partial x^{\eta_{r}}} {\partial X^{\lambda_{r}}}
\mbox{Trpr}\bar{\partial}_{\alpha}\mbox{Trpr}\bar{\partial}_{\beta}
..\mbox{Trpr}\bar{\partial}_{\gamma}{\cal
K}_{\eta_{1}..\eta_{r}}\,,
\end{equation}
where the transverse projection operator is defined by
\begin{equation} \left(\mbox{Trpr}
\,{\cal
K}\right)_{\lambda_{1}..\lambda_{r}}\equiv\theta^{\eta_{1}}_{\lambda_{1}}
..\theta^{\eta_{l}}_{\lambda_{r}}{\cal
 K}_{\eta_{1}..\eta_{r}}\,.\nonumber
\end{equation}
 Actually, our main task will be to transcribe the action of
the covariant derivatives $\nabla_{\mu}$. In particular the resulting expression for the intrinsic d'Alembertian operator (in local coordinates)
$\Box_{H}=\nabla_{\mu}\nabla^{\mu}$ will be very useful since the d'Alembertian appears in every field equation. The transcription of the action of $\Box_{H}$ upon ambient space fields depends on the rank of the tensor used to represent the field.

For a scalar field $\phi$, the situation is simple since $\Box_H$ is linked to the scalar part $Q_{0}^{(1)}$ of the Casimir operator $Q_{}^{(1)}$ through
$$\Box_H \phi(X) =\bar\partial^2\phi(x)=-H^{2}Q_{0}^{(1)}\phi(x)\,.$$

For a  $r$-rank tensor field, the corresponding expression  depends upon $r$ and  contains the operator $Q^{(1)}_{0}$. In Appendix A we prove that
\begin{eqnarray}\label{pass}
\Box_{H}h_{\mu_{1}..\mu_{r}}(X)&=&\frac{\partial
x^{\beta_{1}}}{\partial X^{\mu_{1}}}.. \frac{\partial
x^{\beta_{r}}} {\partial X^{\mu_{r}}}\Big{[}-H^{2}\left(
Q^{(1)}_{0}+r\right){\cal K}_{\beta_{1}..\beta_{r}}
+2H^{4}\sum_{j=1}^{r}x_{\beta_{j}}\sum_{i<j}^{}x_{\beta_{i}} {\cal
K
}_{\beta_{1}..\hat{\beta_{i}}\hat{\beta_{j}}..\beta_{r}}'\nonumber\\
&-&2H^{2}\sum_{i=1}^{r}x_{\beta_{i}}\left( \bar\partial\cdot{\cal
K }_{\beta_{1}..\hat{\beta_{i}}..\beta_{r}}-H^{2}x\cdot{\cal K
}_{\beta_{1}..\hat{\beta_{i}}..\beta_{r}} \right)\Big{]}\,,
\end{eqnarray}
where the symbol $\hat{\beta_{i}}$ indicates that this index
should be removed, and where  ${\cal K}'$ is the trace of ${\cal
K}$ defined by
\begin{equation}
{\cal K}'\equiv{\cal
K}_{\alpha_1..\alpha_{r-2}}=\eta^{\alpha_{r-1}\alpha_{r}}{\cal
K}_{\alpha_1..\alpha_{r}}\,.
\end{equation}
Note that the trace of a tensor of rank r is a tensor of rank r-2.

Now, in the following, we present two
important relations which make explicit the link between the operators $Q^{(1)}_{0}$  and  $Q^{(1)}_{}$ for integer and half-integer spin fields. Thus, with
the help of equation ($\ref{pass}$), any intrinsic  field equation
which contains a d'Alembertian operator and covariant derivatives
will be expressed in terms of $Q^{(1)}_{}$. In this way,  we will be able to
recast any intrinsic field equation in a form comparable to Eq.(\ref{eq:casimir2}).
\subsection{Integer spin  case}

The Casimir operator are simple to manipulate in ambient space
notations. In particular it is easy to show that for a $r$-rank
tensor ${\cal K}_{\eta_{1}..\eta_{r}}(x)$ one has
\begin{equation}\label{eq:integer}
Q^{(1)}{\cal K}(x)=\left(Q_{0}^{(1)}-r(r+1)\right){\cal
K}(x)+2\,\eta \,{\cal K}'+2{\cal S}\, x\,\partial\cdot{\cal
K}(x)-2{\cal S}\,\partial\, x\cdot{\cal K}(x)\,,
\end{equation}
where ${\cal K}'$ is the trace of the $r$-rank tensor ${\cal
K}(x)$ viewed as a homogeneous function of the variables
$x^{\alpha}$ and where ${\cal S}$ is the non normalized
symmetrization operator defined for two vectors $\xi_{\alpha}$ and
$\omega_{\beta}$ by ${\cal
S}(\xi_{\alpha}\omega_{\beta})=\xi_{\alpha}\omega_{\beta}+\xi_{\beta}\omega_{\alpha}$.

This indeed follows from \cite{gahamu}
\begin{equation}\label{rel33}
Q^{(1)}{\cal
K}(x)=-\frac{1}{2}L^{(r)}_{\alpha\beta}L^{\alpha\beta(r)}{\cal
K}(x)=-\frac{1}{2}M_{\alpha\beta}M^{\alpha\beta}{\cal K}(x)
-\frac{1}{2} S^{(r)}_{\alpha\beta}S^{\alpha\beta(r)}{\cal K}(x)
-M_{\alpha\beta}S^{\alpha\beta(r)}{\cal K}(x)\,,
\end{equation}
and
\begin{equation}\label{rel34}
\frac{1}{2} S^{(r)}_{\alpha\beta}S^{\alpha\beta(r)}{\cal K
}(x)=r(r+3){\cal K}(x)-2\eta{\cal K}'\,, \qquad
M_{\alpha\beta}S^{\alpha\beta(r)}{\cal K}(x)=2{\cal S}\,\partial
x\cdot {\cal K}(x)-2{\cal S}\,x\,\partial \cdot {\cal K}(x) -2r\,
{\cal K}(x) \,.
\end{equation}
As we have said,   Eq. (\ref{eq:integer}) will be useful to
express  $Q^{(1)}_{0}$ in terms of $Q^{(1)}_{}$ which given
($\ref{pass}$) will allow to write  every intrinsic field equation
in the form of  (\ref{eq:casimir2}). One can distinguish among the
fields between those which satisfy   the properties of
tracelessness and divergenceless and the others.
\begin{itemize}
\item For the massive fields  the tensor can be chosen  to be
traceless and divergenceless. These conditions which must be
consistent with the field equation \cite{hindawi} allow to
constrain the number of propagating  degrees of freedom (dof) to
the $2s+1$ degrees of a massive field.   In our view, these fields
should be associated to the principal series of unitary
representation because of their minkowskian behaviour in the flat
limit. For such fields, the eigenvalue equation becomes
\begin{equation}
\left(Q_{0}^{(1)}-r(r+1)-\langle Q^{(1)}_{} \rangle\right){\cal
K}(x)=0\quad\mbox{with}\quad \langle Q^{(1)}_{}\rangle
\quad\mbox{corresponding to the princ. series of UIR}\,.\nonumber
\end{equation}
Since in that case (traceless, divergenceless), one has
\begin{equation}
\Box_{H}h_{\mu_{1}..\mu_{r}}(X)=-\frac{\partial
x^{\alpha_{1}}}{\partial X^{\mu_{1}}} ..\frac{\partial
x^{\alpha_{r}}} {\partial
X^{\mu_{r}}}\left(H^{2}Q_{0}^{(1)}+H^{2}r\right){\cal
K}_{\alpha_{r}..\alpha_{r}}(x)\,,
\end{equation}
it follows that the field equation in local coordinates  writes
\begin{equation}\label{eq:eqchampss}
\left(\Box_{H}+H^{2}r(r+2)+H^{2}\langle
Q^{(1)}_{}\rangle\right)h_{\mu_{1}..\mu_{r}}(X)=0\;.
\end{equation}
This expression explicitly conveys the information contained in
$\langle Q^{(1)}_{}\rangle$ to the tracelesss and divergenceless
field in  intrinsic notations. {\bf This equation really connects (in the massive case)  two approaches  largely developed in the framework of field theory, but sometimes a little  independently: field equations in local coordinates and group representation theory}.

Actually, any massive (in the minkowskian sense) spin $s=r$ field
equation on dS space can be written under the above form. Equation
$(\ref{eq:eqchampss})$ therefore allows to unambiguously identify
any massive field equation on ds space.  Note that we will come
across  fields satisfying Eq. $(\ref{eq:eqchampss})$ but with
$\langle Q^{(1)}_{}\rangle$ different from the principal series.
For instance this will happen for the members of the complementary
series and for  the partially massless fields belonging to the
discrete series.  But as we have said  we do not consider these
fields as true massive fields since they do not admit a massive
minkowskian interpretation in the $H=0$ limit. Moreover, we will
see that although the partially massless fields  can be taken
tracelesss and divergenceless they have additional properties
which aren't expected for a massive field (see below). Therefore
we really distinguish among the fields satisfying
$(\ref{eq:eqchampss})$, the subspace characterized by $\langle
Q^{(1)}_{}\rangle$  in the principal series as the true massive
fields. \item For the UIR's different from the principal series,
it is not always possible to impose the  divergenceless or
traceless conditions (see for instance the massless spin-$1$ and
spin-$2$ fields). One can still rewrite the corresponding field
equation in terms of $Q^{(1)}_{}$ and search for the relevant
physical subspace corresponding to a massless UIR of the dS group.
We have seen that  these UIR's correspond to the lowest end of the
discrete series with $p=q=s$ with Casimir eigenvalue $\langle
Q^{(1)}_{p=q}\rangle=-2(s^2-1)$. Therefore, the field equation
will be of the form
\begin{equation}\label{eq:eqchamps}
\left(\Box_{H}+H^{2}r(r+2)-2H^{2}(s^2-1)\right)h_{\mu_{1}..\mu_{r}}(X)+G(x)=0\;,
\end{equation}
where $G(x)$ depends on divergencies and traces of $h$. Examples
are given by the massless spin-$1$ and spin-$2$ fields where the
physical subspaces respectively correspond to the field equations
\cite{allen1,anilto2}
$$\left(\Box_{H}+3H^{2}\right)h_{\mu_{1}}(X)
\qquad\mbox{and}\qquad\left(\Box_{H}+2H^{2}\right)h_{\mu_{1}\mu_{2}}(X).$$

\end{itemize}
\subsection{Half-integer spin  case}
In that case one can represent the field with a four component
symmetric tensor spinor  ${\cal K}(x)={\cal
K}^{i}_{\eta_{1}..\eta_{r}}(x)$ with $i=1,..,4$. The difference
with the integer case is due to the action of the spinorial part
${\cal S }_{\alpha\beta}$ which now reads $${\cal
S}_{\alpha\beta}^{(s)}={\cal S }_{\alpha\beta}^{(r)}+{\cal S
}_{\alpha\beta}^{(\frac{1}{2})}\qquad\mbox{with}\qquad {\cal S
}_{\alpha\beta}^{(\frac{1}{2})}=-\frac{i}{4}[\gamma_{\alpha},\gamma_{\beta}]\qquad\mbox{and}\qquad
s=r+\frac{1}{2}\,,
$$ and with the  Dirac gamma
matrices
$$\gamma^{0}=\left(\begin{array}{cc} \1 &
0\\0&-\1\end{array}\right)\;,\gamma^{1}=\left(\begin{array}{cc} 0
&
i\sigma^{1}\\i\sigma^{1}&0\end{array}\right)\;,\gamma^{2}=\left(\begin{array}{cc}
0 &
-i\sigma^{2}\\-i\sigma^{2}&0\end{array}\right)\;,\gamma^{3}=\left(\begin{array}{cc}
0 &
i\sigma^{3}\\i\sigma^{3}&0\end{array}\right)\;,\gamma^{4}=\left(\begin{array}{cc}
0 & \1\\-\1&0\end{array}\right)\;.$$ Note that ${\cal
S}^{\frac{1}{2}}_{\alpha\beta}$ acts only upon the index $i$. Now
using \cite{lesimple}
$${\cal S}^{(\frac{1}{2})}_{\alpha\beta}{\cal S}^{\alpha\beta(r)}{\cal K}(x)=r {\cal K}(x)-{\cal S}\gamma(\gamma\cdot
{\cal K}(x))\,,$$ and the above relations ($\ref{rel33}$) and
($\ref{rel34}$) one finaly obtains
\begin{equation}\label{eq:seminteger}
Q^{(1)}{\cal K}(x)=\left(Q_{0}^{(1)}-r(r+2)\right){\cal
K}(x)+2\,\eta \,{\cal K}'+2{\cal S}\, x\,\partial\cdot{\cal
K}(x)-2{\cal S}\,\partial\, x\cdot{\cal
K}(x)+\left(\frac{i}{2}\gamma_{\alpha}\gamma_{\beta}M^{\alpha\beta}-\frac{5}{2}\right){\cal
K }+{\cal S}\gamma\gamma\cdot{\cal K}(x)\,.\nonumber
\end{equation}
  Again it is possible with
help of this relation to express the intrinsic wave equation in
terms of the Casimir operator $Q_{}^{(1)}$.

\setcounter{equation}{0}
\section{The mass relation}

We  would now like to introduce a  parameter $m^{2}_{H}$ for dS
fields. Since after all it labels the   UIR's, we at least expect
it to be complete (every UIR labelled). Thus, the massless cases
must correspond to the lowest values of the Casimir operators for
a given spin (the discrete series of representations because of
their minkowskian massless interpretation).  Moreover, we do not
want to systematically interpret any parameter as a mass since we
have seen that not every UIR contracts toward the well known
Poincar\'e parameters. In order to understand  this difference we
must be able to relate the mass parameter to these UIR's.  In
short, given that we have taken care not to introduce in the field
equation any arbitrary parameter, we do not want to do it at this
stage! This is precisely what we will do: the mass parameter will
be unambiguously defined with reference to the UIR's. First it
should be noted that in our classification all the UIR's which
admit a minkowskian interpretation have in common that the spin is
given by the value of $p$. From that, we would like to fix the
zero of the mass parameter for a given $p$. A consistent way of
introducing a positive mass parameter $m^{2}_{H}$ for a given
spin, is to connect the mass parameter to the UIR' s of the dS
group in such a way that the UIR which corresponds to the strictly
massless field yields the value zero for the parameter
$m^{2}_{H}$. Of course the masslessness of these UIR's must be
understood in a minkowskian sense, which means that these UIR's
are precisely those which admit a clear minkowskian masslessness
and spin $s$ interpretation. Our strategy is now clear: for each
value of $p$ we search for the UIR corresponding to the
minkowskian massless field and set the zero of $m^{2}_{H}$  in
reference to that UIR.
\subsection{ Non zero spin fields }

In that case, the UIR's which admit a minkowskian massless
interpretation are located at the bottom of the discrete series
(with $p=q=s$ ) for $s>0$. It can be checked that for a given
$p>0$ these UIR's also correspond to the lowest values of the
Casimir operator!  We are now in position to define a consistent
``mass" parameter with respect to the UIR's characterized by the
values $p=q=s$ (and where $\langle Q_{p=q}^{(1)}\rangle$ is the
corresponding value of the Casimir) by
\begin{equation}\label{mass1}
m_{H}^{2}= H^{2}\left(\langle Q_{}^{(1)}\rangle-\langle
Q_{p=q}^{(1)}\rangle  \right) =\langle Q_{}^{(1)}\rangle
H^2+2(p^2-1)H^2 =\left[(p-q)(p+q-1)\right]H^2\,.
\end{equation}

Since we have set the zero of our mass parameter according to the lowest value of the Casimir operator, this ensures that every UIR for that spin are
labelled by $m^{2}_{H}$ and with $m^2_H\geq ò0$.

We insist, that this parameter is a true mass only
when $\langle Q^{(1)}\rangle $ belongs to the principal series of
unitary representation or to the mentioned massless UIR's. In
these cases the parameter $p$ also corresponds to the spin and can
therefore be replaced by s . For example, the field equation for
massive integer spin fields (i.e princ. series of UIR) given by
(\ref{eq:eqchamps}) finally takes the form
\begin{equation}\label{eq:eqchampsm}
\left(\Box_{H}+\left[2-s(s-2)\right]H^{2}
+m_{H}^{2}\right)h_{\mu_{1}..\mu_{s}}(X)=0\;.
\end{equation}
Whenever $\langle Q^{(1)}\rangle $ does not belong to a UIR with
possible minkowskian interpretation,  we can still use $m^{2}_{H}$
but without referring to a minkowskian mass meaning.

\subsection{ The scalar fields}

We have seen previously that in the scalar case, the massless UIR (with the conformal invariance) corresponds  to the complementary series with the values $p=0,\;q=1$. But  this UIR is weyl equivalent (same eigenvalue for the Casimir operator) to the UIR with values $\Delta=(p,1-q)$, in that case $\Delta=(0,0)$. Hence we can also define the scalar mass relation with respect to the UIR characterized by $p=q$ and denoted by $\langle Q_{p=q}^{(1)}\rangle$. The mass relation given by Eq. (\ref{mass1}) is therefore valid also in the scalar case.

Finally let us mention some peculiarities of the scalar case.
\begin{itemize}
\item [$\bullet$]

Note that by setting the massless
case in reference to the conformally invariant UIR, we implicitly
have assumed the coupling $\xi=1/6$ between the field and  the
background in the usual scalar wave equation
\begin{equation}\label{www}
\left(\Box_{H}+m^{2}_{H}+\xi R\right)\phi=0\,,
\end{equation}
where $\xi$ is the coupling constant and $R=12H^2$ the Ricci
scalar.

\item[ $\bullet$]
Note  that our mass relation given by
\begin{equation}\label{mass11}
m_{H}^{2} =\left[(p-q)(p+q-1)\right]H^2\,,
\end{equation}
also describes the massless   minimally coupled
scalar field.  This field
corresponds to the values  $m^2_{H}=\xi=0$ in the  wave equation (\ref{www}). The involved UIR belongs to the discrete series of UIR $\Pi_{1,0}$ with the values $p=1,\,q=0$ and the eigenvalue $\langle Q_{}^{(1)}\rangle=0$. It is straightforward to verify that $m_H^2=0$ for $p=1, q=0$ in the mass definition (\ref{mass11}).  This field is said to be massless but since the UIR $\Pi_{1,0}$ cannot be linked to the Poincar\'e group,  $m_H$ does not represent a mass and the value $p=1$ does not represent the spin.
\item[ $\bullet$]

In the scalar case,  the UIR $\langle Q_{p=q=0}^{(1)}\rangle=2$ is
not the lowest possible eigenvalue for $p=0$. In fact the
eigenvalues of the Casimir operator with $p=0$ and $-1<q<0$ (also
their weyl equivalent UIR's with $1<q<2$) are smaller. This also
authorizes imaginary values of $m_H$. This is not a serious
inconvenient since on the one hand  the corresponding UIR's have
no minkowskian counterpart and on the other hand these are the
only UIR's which do not yield $m_H^2\in\setR^+$.

\end{itemize}

\subsection{Discussion}
First let us state that one can use the mass relation (\ref{mass1}) even if the
involved UIR are not linked to the Poincar\'{e} group through
group representation contraction. This corresponds to the fact that we do not expect $m^2_H$ to be interpretable as a minkowskian mass for every couple $(p,q)$.

It is easy to check that $m^{2}_{H}$ defined through the equation
($\ref{mass1}$), is a real number for every UIR listed in the
above classification. This  actually is  a consequence of the fact
that the Casimir eigenvalue are real valued. Indeed, the  only
case where a complex number appears in $\langle Q_{}^{(1)}\rangle$
concerns the principal series with $q=1/2+i\nu$ and $\nu\in\setR$.
Since it occurs under the form $(q+1)(q-2)$ it also yields a real
$m^{2}_{H}$. In fact {\bf the value of $m^{2}_{H}$ is a  positive
real number} (it is constructed in that way) except for the UIR's
with $p=0$ and $-1<q<0,\; 1<q<2$ but which as we have said do not
correspond to any minkowskian UIR in the $H=0$ limit.

Let us finally examine the vanishing curvature limit ($H=0$). One
distinguishes essentially two cases. If  the mass parameter is
defined with $\langle Q_{}^{(1)}\rangle$ belonging to the
principal series of UIR, then its limit is well defined. In this
case $p$ remains constant, and the hole family of UIR's with
$0<q<+\infty$ contract toward the massive Poincar\'{e} group UIR.
The parameter $q$ is then given by  $q\sim-i m/H$ such that
$m^{2}_{H}\sim m $ where m is the minkowskian mass. On the other
hand, for all the representations different from the principal
series and which yield a non zero mass parameter, the limit gives
$m^{2}_{H}=0$. This is because in the complementary and discrete
series of representations a fixed value of  $p$ entails a bounded
value of $q$. Therefore there is no term in $m^2_{H}$ to balance
the vanishing $H$. Note that this vanishing mass parameter should
not be interpreted as a minkowskian massless field since at the
level of group representation the limit is not well defined.
Although the group contraction procedure allows to give an insight
into the desitterian physics, the status of the purely desitterian
field is not really understood. Because these fields can still be
considered at very large radius (locally almost flat) we do not
know how a minkowskian observer would perceive them. Certainly not
in terms of mass and spin.

With the help of examples we now show how our properly defined
mass relations  really simplifies the issues of recent debates.
These are first the new gauge states known as ``partially
massless" fields largely discussed in a recent series of papers of
S. Deser and A. Waldron in \cite{deser2,deser3,deser4,deser5}.
Secondly we  address  the question of the mass of the graviton in
(A)dS space which has recently been debated by M. Novello and R.
P. Neves in \cite{novello1}.

\setcounter{equation}{0}
\section{Examples and applications}
We can now use our properly defined mass definition to label the
fields  in dS space according to $(m_H^2,p=s)$. This will be
useful since a large amount of authors actually use a mass
parameter in their field equation. Our main advantage is that not
only we will be able to trace back the involved UIR for a given
value of $m_H^2$ (and thus check if the corresponding mass
deserves the mass denomination by considering the UIR
contraction), but we are also sure that every UIR of the above
classification can be described by $m_H^2$. We will illustrate the
efficiency of our approach with two examples.
\subsection{Gauge invariant  fields and group representation}

First of all, we examine the mass values of a given type of fields
characterized by gauge invariance. In fact, we will call
(strictly) massive a field that has $2s+1$ degrees of freedom
(dof) with all the $2s+1$ helicities and no gauge invariance.
Starting from there  one distinguishes essentially two classes of
{\bf gauge invariant} fields:
\begin{enumerate}
\item Strictly massless fields where the gauge invariance allows
to reduce the degrees of freedom to two (helicities $\pm s$). The
corresponding field equation are invariant under conformal
transformation. \item Intermediate fields where the gauge
invariance allows to remove subsets of lowest helicity modes. Such
fields which correspond to a novel gauge invariance, were first
observed by S. Deser in \cite{deser1} in the spin $2$ case.
Following S. Deser and A. Waldron \cite{deser3,deser4,deser5} we
will designate these fields  as partially massless fields.  This
designation is due to their light cone propagation properties
\cite{deser6}. The new gauge invariance permits ``intermediate
sets of higher helicities " between the $2s+1$  massive degrees of
freedom (dof) and the $2$ strictly massless helicities.
\end{enumerate}

Of course, all the mentioned  fields can be  completely
characterized by their $(m_H^2,s)$ values. Given a value of
$s>1$, S. Deser and A. Waldron \cite{deser3,deser4,deser5,deser6}
have shown   (for de Sitter or anti-de Sitter) that the plane
defined by a mass parameter $m_{H}^{2}$   and the cosmological
constant $\Lambda=\pm 3H^2$ was divided into different phases
which correspond to unitary or non-unitary regions. The non
unitary regions correspond to forbidden mass ranges. In the
spin-$2$ case for instance,  this mass range   has been  discussed
by Higuchi in \cite{higu2}. These regions are separated by lines
of the  gauge invariant fields described above.

In the following we examine the mass values of the gauge invariant
fields in the de Sitter case ($\Lambda=3H^2$). Beyond the mere
value of $m_H^2$, our  mass definition  linked to the UIR's of the
de Sitter group enables us to characterize  the partially massless
fields following their group representation content. Actually, we
would like to show how the partially massless fields with spin $s$
are linked to very specific   representations of the dS group and
how the forbidden mass ranges correspond to the gaps between two
unitary representations in the classification. This is easily done
if we compare for a given spin, the mass relation given by
(\ref{mass1}) with the ($m_{\sss DW}^{2},\Lambda$) pictures found
in \cite{deser4} where $m_{\sss DW}$ designates the mass used by
S. Deser and A. Waldron. Since the authors of \cite{deser6}
consider both dS and AdS geometries, their figures include
negative values of the cosmological constant. But as our mass
relation is given in terms of the  UIR's of the dS group, we
restrict ourself to the unitary  regions where $\Lambda>0$.

For integer spins our mass definition (\ref{mass1}) agrees with
theirs, in particular the strictly massless case (which we have
used to set the zero of $m_H^2$) yields the value $m_{\sss
DW}^2=0$.  Therefore we set $m_{\sss DW}^{2}=m^{2}_{H}$ in the
following for the integer case.

The situation is a little different for half integer spins. In
\cite{deser4} it is argued that the partially massless fields
correspond to AdS fields (in order to have a positive mass, see
Figure \ref{f5}) because otherwise it would entail negative values
of $m_{\sss DW}^2$. On the contrary we believe that these  fields
actually can be found in de Sitter space  and that the negative
mass values are merely a consequence of a bad choice of mass
relation. For instance, we will see that with the choice of mass
$m_{\sss DW}^2$, {\bf the strictly massless fields have a mass
$m_{\sss DW}^2\neq 0$  whereas  they would  yield  $m_H^2=0$ if
defined according to (\ref{mass1})}.

Finally for integer and half integer spin fields, we will see that
{\bf all the gauge invariant fields correspond to the family of
the discrete series of representation} of the de Sitter group.

\vspace{0.3cm}\noindent{\underline{\bf Integer spins examples}
}\vspace{0.3cm}

As announced,  the  mass definition (\ref{mass1}) agrees with that
given in \cite{deser4}.  In the following we examine the
$(m_H^2,\Lambda=3H^2)$ figures for spins up to s=3 which are given
in \cite{deser4}. We would like to see how the gauge fields can be
characterized.
\begin{itemize}
\item For the scalar and vector cases, there are no positive
forbidden mass ranges (no new gauge invariance occurs). This can
be seen on Figure \ref{f1} since for $p=0$ and $p=1$ one covers the
unitary region continuously. The corresponding mass relations are
shown in Figure \ref{f2}. (the discrete series members are represented
by the symbol $\square$ ).
\begin{figure}[h]
\begin{center}

\begin{picture}(0,0)%
\includegraphics{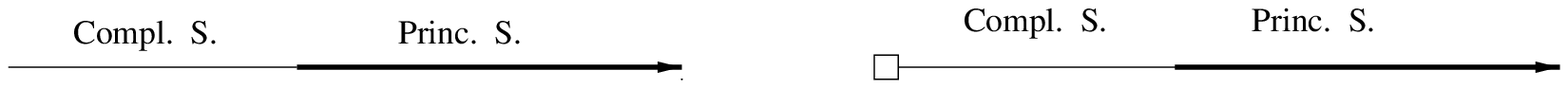}%
\end{picture}%
\setlength{\unitlength}{3197sp}%
\begingroup\makeatletter\ifx\SetFigFont\undefined
\def\x#1#2#3#4#5#6#7\relax{\def\x{#1#2#3#4#5#6}}%
\expandafter\x\fmtname xxxxxx\relax \def\y{splain}%
\ifx\x\y   
\gdef\SetFigFont#1#2#3{%
  \ifnum #1<17\tiny\else \ifnum #1<20\small\else
  \ifnum #1<24\normalsize\else \ifnum #1<29\large\else
  \ifnum #1<34\Large\else \ifnum #1<41\LARGE\else
     \huge\fi\fi\fi\fi\fi\fi
  \csname #3\endcsname}%
\else \gdef\SetFigFont#1#2#3{\begingroup
  \count@#1\relax \ifnum 25<\count@\count@25\fi
  \def\x{\endgroup\@setsize\SetFigFont{#2pt}}%
  \expandafter\x
    \csname \romannumeral\the\count@ pt\expandafter\endcsname
    \csname @\romannumeral\the\count@ pt\endcsname
  \csname #3\endcsname}%
\fi \fi\endgroup
\begin{picture}(9933,1143)(376,-2986)
\put(4576,-2461){\makebox(0,0)[lb]{\smash{\SetFigFont{11}{13.2}{rm}$m_{H}^{2}$}}}
\put(7351,-2011){\makebox(0,0)[lb]{\smash{\SetFigFont{11}{13.2}{rm}Spin
$1$}}}
\put(1876,-2011){\makebox(0,0)[lb]{\smash{\SetFigFont{11}{13.2}{rm}Spin
$0$}}}
\put(376,-2986){\makebox(0,0)[lb]{\smash{\SetFigFont{11}{13.2}{rm}$0$}}}
\put(9976,-2461){\makebox(0,0)[lb]{\smash{\SetFigFont{11}{13.2}{rm}$m_{H}^{2}$}}}
\put(7801,-2986){\makebox(0,0)[lb]{\smash{\SetFigFont{11}{13.2}{rm}$H^2/4$}}}
\put(6001,-2986){\makebox(0,0)[lb]{\smash{\SetFigFont{11}{13.2}{rm}$0$}}}
\put(2251,-2986){\makebox(0,0)[lb]{\smash{\SetFigFont{11}{13.2}{rm}$H^2/4$}}}
\end{picture}

\caption{Scalar  and
vector ($p=0$ and $p=1$) mass relations.}\label{f2}
\end{center}
\end{figure}
\item According to \cite{deser6}  the spin-$2$ partial massless
gauge field is given by the value $m_{H}^2=2\Lambda/3=2H^{2}$ and
the strictly massless field corresponds to  $m_{H}^2=0$. One
notices with the help of definition (\ref{mass1}) that both belong
to the discrete series of unitary irreducible representations
corresponding to the values $p=2$ with $q=1$ ($m_{H}^2=2H^{2}$) or
$q=2$ ($m_{H}^2=0$). Both cases are characterized by a certain
gauge invariance which allows to reduce the number of the degrees
of freedom of the corresponding fields. One gets two helicities in
the case $m_{H}^2=0$ and four degrees of freedom for the field
with $m_{H}^2=2H^{2}$. The strictly massless case $m_{H}^2=0$ with
$p=q=2$ corresponds to the spin-$2$ linearized gravity (see Figure
\ref{f3}).
\begin{figure}[h]
\begin{center}

\begin{picture}(0,0)%
\includegraphics{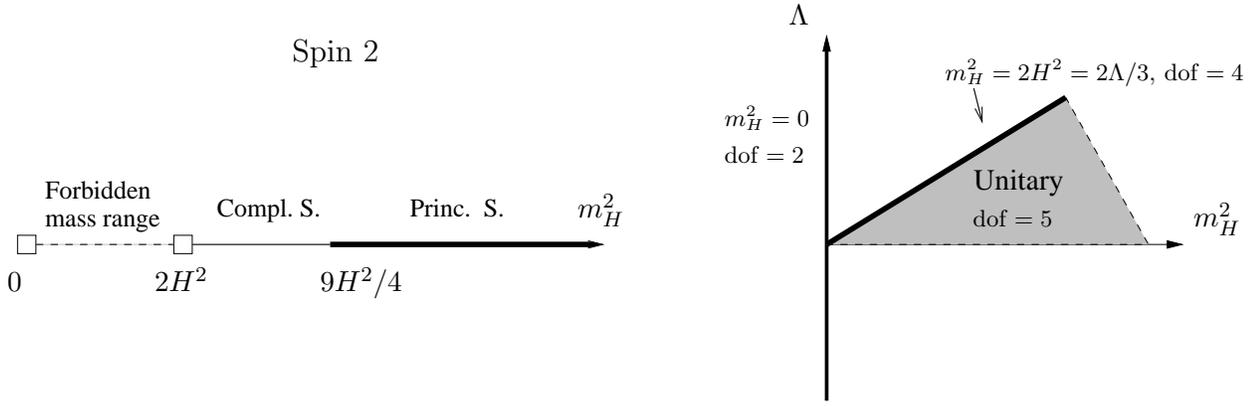}%
\end{picture}%
\setlength{\unitlength}{3039sp}%
\begingroup\makeatletter\ifx\SetFigFont\undefined
\def\x#1#2#3#4#5#6#7\relax{\def\x{#1#2#3#4#5#6}}%
\expandafter\x\fmtname xxxxxx\relax \def\y{splain}%
\ifx\x\y   
\gdef\SetFigFont#1#2#3{%
  \ifnum #1<17\tiny\else \ifnum #1<20\small\else
  \ifnum #1<24\normalsize\else \ifnum #1<29\large\else
  \ifnum #1<34\Large\else \ifnum #1<41\LARGE\else
     \huge\fi\fi\fi\fi\fi\fi
  \csname #3\endcsname}%
\else \gdef\SetFigFont#1#2#3{\begingroup
  \count@#1\relax \ifnum 25<\count@\count@25\fi
  \def\x{\endgroup\@setsize\SetFigFont{#2pt}}%
  \expandafter\x
    \csname \romannumeral\the\count@ pt\expandafter\endcsname
    \csname @\romannumeral\the\count@ pt\endcsname
  \csname #3\endcsname}%
\fi \fi\endgroup
\begin{picture}(10650,3288)(376,-4069)
\put(376,-3136){\makebox(0,0)[lb]{\smash{\SetFigFont{11}{13.2}{rm}$0$}}}
\put(2926,-3136){\makebox(0,0)[lb]{\smash{\SetFigFont{11}{13.2}{rm}$9H^{2}/4$}}}
\put(10051,-2611){\makebox(0,0)[lb]{\smash{\SetFigFont{11}{13.2}{rm}$m_{H}^{2}$}}}
\put(11026,-1411){\makebox(0,0)[lb]{\smash{\SetFigFont{9}{10.8}{rm}
}}}
\put(6751,-961){\makebox(0,0)[lb]{\smash{\SetFigFont{11}{13.2}{rm}$\Lambda$}}}
\put(6226,-1786){\makebox(0,0)[lb]{\smash{\SetFigFont{9}{10.8}{rm}$m_{H}^{2}=0$
}}}
\put(6226,-2086){\makebox(0,0)[lb]{\smash{\SetFigFont{9}{10.8}{rm}dof
$=2$}}}
\put(2701,-1261){\makebox(0,0)[lb]{\smash{\SetFigFont{12}{14.4}{rm}Spin
$2$}}}
\put(8026,-1411){\makebox(0,0)[lb]{\smash{\SetFigFont{9}{10.8}{rm}$m_{H}^{2}=2H^{2}=2\Lambda/3,$
dof $=4$}}}
\put(8251,-2611){\makebox(0,0)[lb]{\smash{\SetFigFont{9}{10.8}{rm}dof
$=5$}}}
\put(1576,-3136){\makebox(0,0)[lb]{\smash{\SetFigFont{11}{13.2}{rm}$2H^{2}$}}}
\put(5026,-2536){\makebox(0,0)[lb]{\smash{\SetFigFont{11}{13.2}{rm}$m_{H}^{2}$}}}
\end{picture}

\caption{Spin-$2$ mass relation and phase diagram showing
partially massless lines.}\label{f3}
\end{center}
\end{figure}

\item For the spin-$3$ (see Figure \ref{f4}) case the lines of new
gauge invariance are given by the values
$m_{H}^2=4\Lambda/3=4H^{2}$, $m_{H}^2=2\Lambda=6H^{2}$ and  the
strictly massless field corresponds to $m_{H}^2=0$. According to
(\ref{mass1}) they belong to the discrete series of unitary
irreducible representations with the values $p=3$ with $q=1$
($m_{H}^2=6H^{2}$), $q=2$ ($m_{H}^2=4H^2$) and $q=3$
($m_{H}^2=0$).
\begin{figure}[h]
\begin{center}

\begin{picture}(0,0)%
\includegraphics{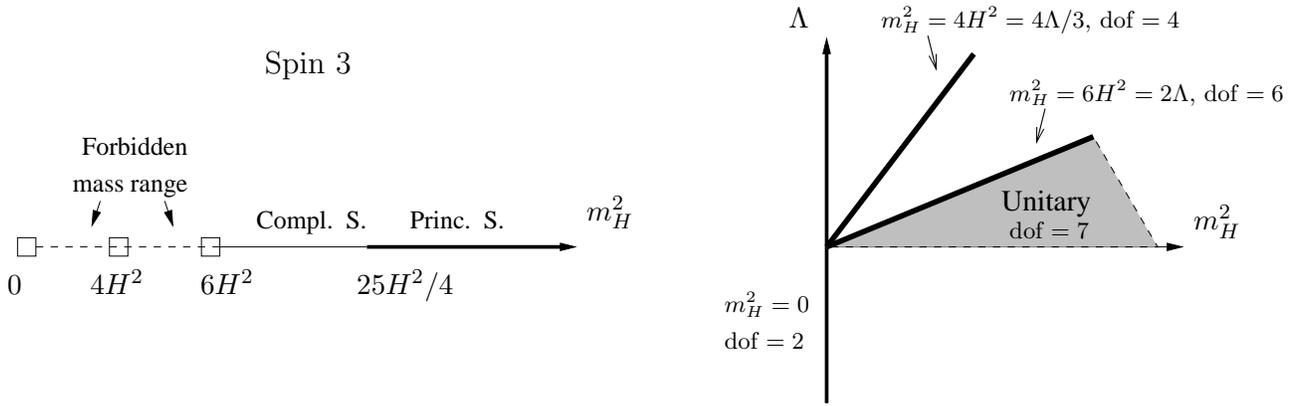}%
\end{picture}%
\setlength{\unitlength}{3039sp}%
\begingroup\makeatletter\ifx\SetFigFont\undefined
\def\x#1#2#3#4#5#6#7\relax{\def\x{#1#2#3#4#5#6}}%
\expandafter\x\fmtname xxxxxx\relax \def\y{splain}%
\ifx\x\y   
\gdef\SetFigFont#1#2#3{%
  \ifnum #1<17\tiny\else \ifnum #1<20\small\else
  \ifnum #1<24\normalsize\else \ifnum #1<29\large\else
  \ifnum #1<34\Large\else \ifnum #1<41\LARGE\else
     \huge\fi\fi\fi\fi\fi\fi
  \csname #3\endcsname}%
\else \gdef\SetFigFont#1#2#3{\begingroup
  \count@#1\relax \ifnum 25<\count@\count@25\fi
  \def\x{\endgroup\@setsize\SetFigFont{#2pt}}%
  \expandafter\x
    \csname \romannumeral\the\count@ pt\expandafter\endcsname
    \csname @\romannumeral\the\count@ pt\endcsname
  \csname #3\endcsname}%
\fi \fi\endgroup
\begin{picture}(10650,3303)(376,-4069)
\put(5101,-2536){\makebox(0,0)[lb]{\smash{\SetFigFont{11}{13.2}{rm}$m_{H}^{2}$}}}
\put(1051,-3136){\makebox(0,0)[lb]{\smash{\SetFigFont{11}{13.2}{rm}$4H^{2}$}}}
\put(10051,-2611){\makebox(0,0)[lb]{\smash{\SetFigFont{11}{13.2}{rm}$m_{H}^{2}$}}}
\put(11026,-1411){\makebox(0,0)[lb]{\smash{\SetFigFont{9}{10.8}{rm}
}}}
\put(6751,-961){\makebox(0,0)[lb]{\smash{\SetFigFont{11}{13.2}{rm}$\Lambda$}}}
\put(6226,-3286){\makebox(0,0)[lb]{\smash{\SetFigFont{9}{10.8}{rm}$m_{H}^{2}=0$
}}}
\put(6226,-3586){\makebox(0,0)[lb]{\smash{\SetFigFont{9}{10.8}{rm}dof
$=2$}}}
\put(376,-3136){\makebox(0,0)[lb]{\smash{\SetFigFont{11}{13.2}{rm}$0$}}}
\put(3226,-3136){\makebox(0,0)[lb]{\smash{\SetFigFont{11}{13.2}{rm}$25H^2/4$}}}
\put(2476,-1336){\makebox(0,0)[lb]{\smash{\SetFigFont{12}{14.4}{rm}Spin
$3$}}}
\put(8551,-2686){\makebox(0,0)[lb]{\smash{\SetFigFont{9}{10.8}{rm}dof
$=7$}}}
\put(8551,-1561){\makebox(0,0)[lb]{\smash{\SetFigFont{9}{10.8}{rm}$m_{H}^{2}=6H^{2}=2\Lambda,$
dof $=6$}}}
\put(7501,-961){\makebox(0,0)[lb]{\smash{\SetFigFont{9}{10.8}{rm}$m_{H}^{2}=4H^{2}=4\Lambda/3,$
dof $=4$}}}
\put(1951,-3136){\makebox(0,0)[lb]{\smash{\SetFigFont{11}{13.2}{rm}$6H^{2}$}}}
\end{picture}

\caption{Spin-$3$ mass relation and phase diagram showing
partially massless lines.}\label{f4}
\end{center}
\end{figure}
\end{itemize}
The situation is  quite clear, the partially massless  and the
strictly massless fields (all characterized by some gauge
invariance) correspond to the members of the discrete series of
representations! They are represented in figure $9$ (the squares
up to spin $3$) where dashed lines indicate the degrees of freedom
shared by the members of the same diagonal. Such a figure provides
an efficient tool for identifying the partially massless lines and
given the mass relation (\ref{mass1}) allows to locate the regions
which correspond to forbidden mass values.

\vspace{0.3cm}\noindent{\underline{\bf Half integer spins examples
}}\vspace{0.3cm}

In the  half integer case, for $s=\frac{1}{2}$, there is no
forbidden region (same as spin-$0$ or spin-$1$). More interesting
are the examples $s=\frac{3}{2}$, and $s=\frac{5}{2}$ because in
that case we disagree with the mass definition given in
\cite{deser6}. Let us present both point of views.

\begin{itemize}
\item

Following the figures (see Figure \ref{f5}) found in \cite{deser6}, it
seems that the ``mass" relations are defined relatively to the
first terms of the discrete series of representation (with $q=1/2$
rather than $q=p$) for a given spin. In other words, this means
that the relation (\ref{mass1}) should be replaced by
\begin{equation}\label{mass4}
m_{DW}^{2}= H^{2}\left(\langle Q_{}^{(1)}\rangle  -\langle
Q_{p,q=1/2}^{(1)}\rangle \right)= \frac{\Lambda}{3}\left(\langle
Q_{}^{(1)}\rangle  -\langle Q_{p,q=1/2}^{(1)}\rangle
\right)=\frac{\Lambda}{3}\left[-q(q-1)-\frac{1}{4} \right].
\end{equation}
\begin{figure}[h]
\begin{center}

\begin{picture}(0,0)%
\includegraphics{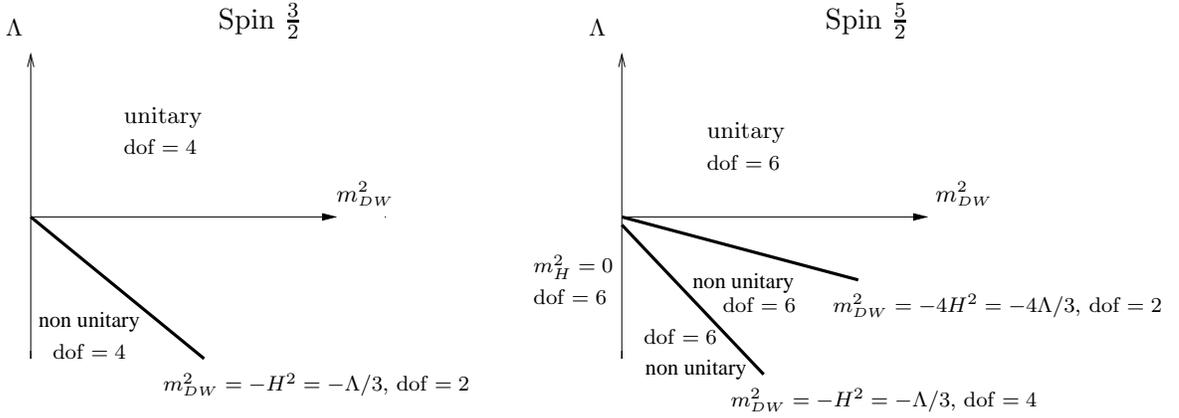}%
\end{picture}%
\setlength{\unitlength}{2605sp}%
\begingroup\makeatletter\ifx\SetFigFont\undefined
\def\x#1#2#3#4#5#6#7\relax{\def\x{#1#2#3#4#5#6}}%
\expandafter\x\fmtname xxxxxx\relax \def\y{splain}%
\ifx\x\y   
\gdef\SetFigFont#1#2#3{%
  \ifnum #1<17\tiny\else \ifnum #1<20\small\else
  \ifnum #1<24\normalsize\else \ifnum #1<29\large\else
  \ifnum #1<34\Large\else \ifnum #1<41\LARGE\else
     \huge\fi\fi\fi\fi\fi\fi
  \csname #3\endcsname}%
\else \gdef\SetFigFont#1#2#3{\begingroup
  \count@#1\relax \ifnum 25<\count@\count@25\fi
  \def\x{\endgroup\@setsize\SetFigFont{#2pt}}%
  \expandafter\x
    \csname \romannumeral\the\count@ pt\expandafter\endcsname
    \csname @\romannumeral\the\count@ pt\endcsname
  \csname #3\endcsname}%
\fi \fi\endgroup
\begin{picture}(9825,3792)(1201,-4561)
\put(11026,-1411){\makebox(0,0)[lb]{\smash{\SetFigFont{8}{9.6}{rm}
}}}
\put(6226,-3286){\makebox(0,0)[lb]{\smash{\SetFigFont{8}{9.6}{rm}$m_{H}^{2}=0$
}}}
\put(7876,-2311){\makebox(0,0)[lb]{\smash{\SetFigFont{8}{9.6}{rm}dof
$=6$}}}
\put(7876,-2011){\makebox(0,0)[lb]{\smash{\SetFigFont{9}{10.8}{rm}unitary}}}
\put(7276,-3961){\makebox(0,0)[lb]{\smash{\SetFigFont{8}{9.6}{rm}dof
$=6$}}}
\put(6226,-3586){\makebox(0,0)[lb]{\smash{\SetFigFont{8}{9.6}{rm}dof
$=6$}}}
\put(4351,-2611){\makebox(0,0)[lb]{\smash{\SetFigFont{9}{10.8}{rm}$m_{\sss
DW}^{2}$}}}
\put(8026,-3661){\makebox(0,0)[lb]{\smash{\SetFigFont{8}{9.6}{rm}dof
$=6$}}}
\put(1651,-4111){\makebox(0,0)[lb]{\smash{\SetFigFont{8}{9.6}{rm}dof
$=4$}}}
\put(1201,-1036){\makebox(0,0)[lb]{\smash{\SetFigFont{9}{10.8}{rm}$\Lambda$}}}
\put(2326,-1861){\makebox(0,0)[lb]{\smash{\SetFigFont{9}{10.8}{rm}unitary}}}
\put(2326,-2161){\makebox(0,0)[lb]{\smash{\SetFigFont{8}{9.6}{rm}dof
$=4$}}}
\put(6751,-1036){\makebox(0,0)[lb]{\smash{\SetFigFont{9}{10.8}{rm}$\Lambda$}}}
\put(3226,-961){\makebox(0,0)[lb]{\smash{\SetFigFont{11}{13.2}{rm}Spin
$\frac{3}{2}$}}}
\put(9001,-961){\makebox(0,0)[lb]{\smash{\SetFigFont{11}{13.2}{rm}Spin
$\frac{5}{2}$}}}
\put(2701,-4411){\makebox(0,0)[lb]{\smash{\SetFigFont{8}{9.6}{rm}$m_{\sss
DW}^{2}=-H^{2}=-\Lambda/3,$ dof $=2$}}}
\put(8101,-4561){\makebox(0,0)[lb]{\smash{\SetFigFont{8}{9.6}{rm}$m_{\sss
DW}^{2}=-H^{2}=-\Lambda/3,$ dof $=4$}}}
\put(9076,-3661){\makebox(0,0)[lb]{\smash{\SetFigFont{8}{9.6}{rm}$m_{\sss
DW}^{2}=-4H^{2}=-4\Lambda/3,$ dof $=2$}}}
\put(10051,-2611){\makebox(0,0)[lb]{\smash{\SetFigFont{9}{10.8}{rm}$m_{\sss
DW}^{2}$}}}
\end{picture}

\caption{Spin-$\frac{3}{2}$ and $\frac{5}{2}$ mass relation
according to (\ref{mass4}).}\label{f5}
\end{center}
\end{figure}

Indeed, following this mass relation, one can check  that the
partially massless lines are disposed for half integer spin fields
as shown in Figure \ref{f5} (therefore in agreement with
\cite{deser6}). The strictly massless fields ($2$ helicities only)
in that case still correspond to the lowest values of the discrete
series of representation namely $\langle Q_{}^{(1)}\rangle$ with
$p=q$ ($m_{\sss DW}^2=-\Lambda/3$ for the spin $\frac{3}{2}$ and
$m_{\sss DW}^2=-4\Lambda/3$ for the spin $\frac{5}{2}$ ). Note also
that the values with $q=1/2$ ($m_{\sss DW}^{2}=0$) are not partially
massless gauge fields. In \cite{deser6} it is claimed that the
gauge lines correspond to AdS fields (since $m^{2}_{\sss DW}<0$ if
$\Lambda>0$), thus there are no strictly massless fields in dS
space for these spin values.
\item
We would like to give some arguments to show why we believe that
this absence of strictly  massless fields in dS space is not a
physical fact but rather that it is due to an erroneous choice of
mass parameter. {\bf First note that the strictly massless fields for
(\ref{mass4}) do not yield the value $m_{\sss DW}^{2}=0$}. Of course one
could say that this is of no importance  since the zero can be
chosen arbitrarily. But having made this choice one should still
take into account the negative values which might occur!
\begin{figure}[h]
\begin{center}

\begin{picture}(0,0)%
\includegraphics{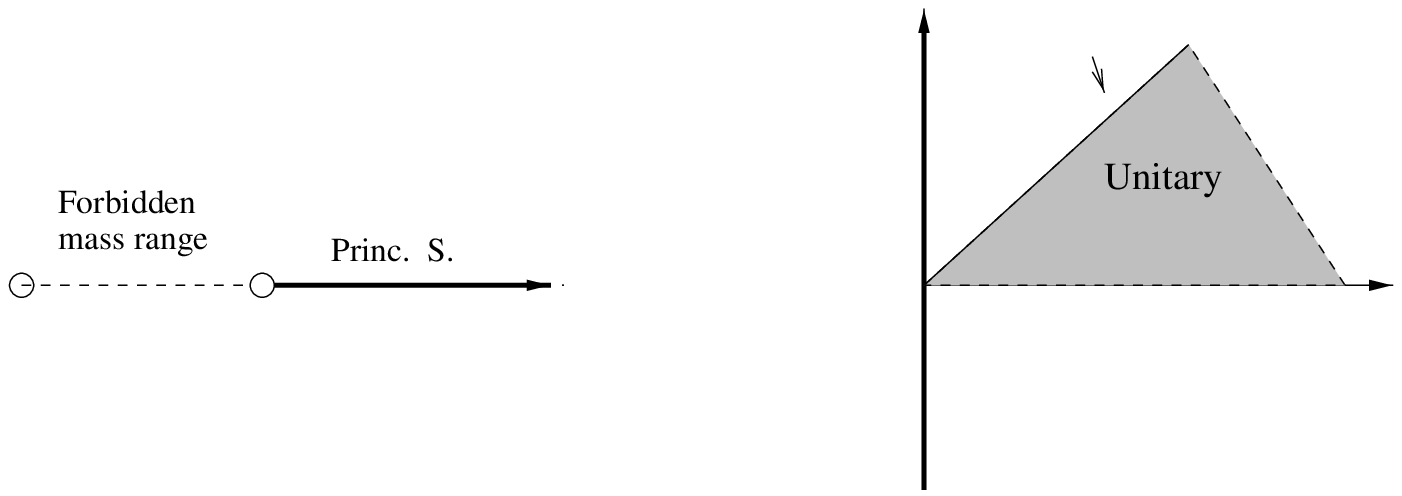}%
\end{picture}%
\setlength{\unitlength}{3039sp}%
\begingroup\makeatletter\ifx\SetFigFont\undefined
\def\x#1#2#3#4#5#6#7\relax{\def\x{#1#2#3#4#5#6}}%
\expandafter\x\fmtname xxxxxx\relax \def\y{splain}%
\ifx\x\y   
\gdef\SetFigFont#1#2#3{%
  \ifnum #1<17\tiny\else \ifnum #1<20\small\else
  \ifnum #1<24\normalsize\else \ifnum #1<29\large\else
  \ifnum #1<34\Large\else \ifnum #1<41\LARGE\else
     \huge\fi\fi\fi\fi\fi\fi
  \csname #3\endcsname}%
\else \gdef\SetFigFont#1#2#3{\begingroup
  \count@#1\relax \ifnum 25<\count@\count@25\fi
  \def\x{\endgroup\@setsize\SetFigFont{#2pt}}%
  \expandafter\x
    \csname \romannumeral\the\count@ pt\expandafter\endcsname
    \csname @\romannumeral\the\count@ pt\endcsname
  \csname #3\endcsname}%
\fi \fi\endgroup
\begin{picture}(9825,3288)(1201,-4069)
\put(11026,-1411){\makebox(0,0)[lb]{\smash{\SetFigFont{9}{10.8}{rm}
}}}
\put(6751,-961){\makebox(0,0)[lb]{\smash{\SetFigFont{11}{13.2}{rm}$\Lambda$}}}
\put(7801,-1186){\makebox(0,0)[lb]{\smash{\SetFigFont{9}{10.8}{rm}$m_{H}^{2}=H^{2}=\Lambda/3,$
dof $=4$}}}
\put(1201,-3136){\makebox(0,0)[lb]{\smash{\SetFigFont{11}{13.2}{rm}$0$}}}
\put(2701,-3136){\makebox(0,0)[lb]{\smash{\SetFigFont{11}{13.2}{rm}$H^{2}$}}}
\put(4651,-2611){\makebox(0,0)[lb]{\smash{\SetFigFont{11}{13.2}{rm}$m_{H}^{2}$}}}
\put(3301,-1411){\makebox(0,0)[lb]{\smash{\SetFigFont{12}{14.4}{rm}Spin
$\frac{3}{2}$}}}
\put(6151,-3061){\makebox(0,0)[lb]{\smash{\SetFigFont{9}{10.8}{rm}dof
$=2$}}}
\put(6151,-2761){\makebox(0,0)[lb]{\smash{\SetFigFont{9}{10.8}{rm}$m_{H}^{2}=0$
}}}
\put(9901,-2536){\makebox(0,0)[lb]{\smash{\SetFigFont{11}{13.2}{rm}$m_{H}^{2}$}}}
\put(8251,-2461){\makebox(0,0)[lb]{\smash{\SetFigFont{9}{10.8}{rm}dof
$=4$}}}
\end{picture}

\caption{Spin-$\frac{3}{2}$ mass relation according to
(\ref{mass1}).}\label{f6}
\end{center}
\end{figure}
In fact  the  consequence of the mass definition (\ref{mass4}) is
that the members  of the discrete series with $p$ fixed and
$q>1/2$ assume negative ``mass" values and as such are therefore
eliminated, although they correspond to unitary representations.
Unfortunately this also eliminates the UIR's $\langle
Q_{}^{(1)}\rangle$ with $p=q$ which is known to correspond to the
strictly massless case!     Actually it is sufficient to adopt the
relation (\ref{mass1}),  in order to have a   non negative mass
for every UIR. Moreover, the strictly massless case is again found
for $p=q$ with $m^{2}_{H}=0$. The resulting $(m^{2}_{H},\Lambda)$
diagrams are represented in the figures \ref{f6} and \ref{f7}.

\begin{figure}[h]
\begin{center}

\begin{picture}(0,0)%
\includegraphics{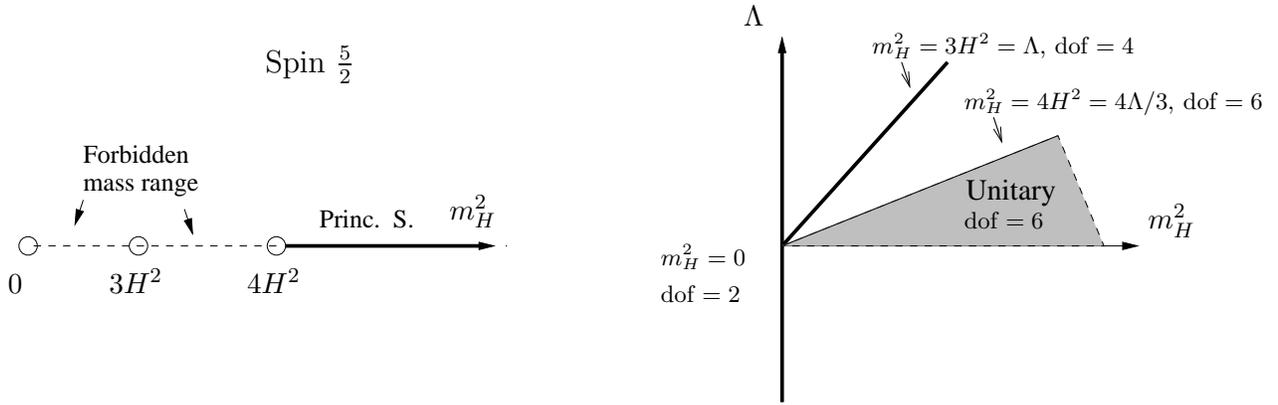}%
\end{picture}%
\setlength{\unitlength}{3039sp}%
\begingroup\makeatletter\ifx\SetFigFont\undefined
\def\x#1#2#3#4#5#6#7\relax{\def\x{#1#2#3#4#5#6}}%
\expandafter\x\fmtname xxxxxx\relax \def\y{splain}%
\ifx\x\y   
\gdef\SetFigFont#1#2#3{%
  \ifnum #1<17\tiny\else \ifnum #1<20\small\else
  \ifnum #1<24\normalsize\else \ifnum #1<29\large\else
  \ifnum #1<34\Large\else \ifnum #1<41\LARGE\else
     \huge\fi\fi\fi\fi\fi\fi
  \csname #3\endcsname}%
\else \gdef\SetFigFont#1#2#3{\begingroup
  \count@#1\relax \ifnum 25<\count@\count@25\fi
  \def\x{\endgroup\@setsize\SetFigFont{#2pt}}%
  \expandafter\x
    \csname \romannumeral\the\count@ pt\expandafter\endcsname
    \csname @\romannumeral\the\count@ pt\endcsname
  \csname #3\endcsname}%
\fi \fi\endgroup
\begin{picture}(10275,3288)(751,-4069)
\put(10051,-2611){\makebox(0,0)[lb]{\smash{\SetFigFont{11}{13.2}{rm}$m_{H}^{2}$}}}
\put(8551,-1636){\makebox(0,0)[lb]{\smash{\SetFigFont{9}{10.8}{rm}$m_{H}^{2}=4H^{2}=4\Lambda/3,$
dof $=6$}}}
\put(11026,-1411){\makebox(0,0)[lb]{\smash{\SetFigFont{9}{10.8}{rm}
}}}
\put(6751,-961){\makebox(0,0)[lb]{\smash{\SetFigFont{11}{13.2}{rm}$\Lambda$}}}
\put(4351,-2536){\makebox(0,0)[lb]{\smash{\SetFigFont{11}{13.2}{rm}$m_{H}^{2}$}}}
\put(2851,-1336){\makebox(0,0)[lb]{\smash{\SetFigFont{12}{14.4}{rm}Spin
$\frac{5}{2}$}}}
\put(751,-3136){\makebox(0,0)[lb]{\smash{\SetFigFont{11}{13.2}{rm}$0$}}}
\put(2701,-3136){\makebox(0,0)[lb]{\smash{\SetFigFont{11}{13.2}{rm}$4H^{2}$}}}
\put(1576,-3136){\makebox(0,0)[lb]{\smash{\SetFigFont{11}{13.2}{rm}$3H^{2}$}}}
\put(6076,-2911){\makebox(0,0)[lb]{\smash{\SetFigFont{9}{10.8}{rm}$m_{H}^{2}=0$
}}}
\put(6076,-3211){\makebox(0,0)[lb]{\smash{\SetFigFont{9}{10.8}{rm}dof
$=2$}}}
\put(8551,-2611){\makebox(0,0)[lb]{\smash{\SetFigFont{9}{10.8}{rm}dof
$=6$}}}
\put(7801,-1186){\makebox(0,0)[lb]{\smash{\SetFigFont{9}{10.8}{rm}$m_{H}^{2}=3H^{2}=\Lambda,$
dof $=4$}}}
\end{picture}

\caption{Spin-$\frac{5}{2}$ mass relation according to
(\ref{mass1}).}\label{f7}
\end{center}
\end{figure}
\end{itemize}

\vspace{0.3cm}\noindent{\underline{\bf Discussion }}\vspace{0.3cm}

\begin{figure}[h]
\begin{center}

\begin{picture}(0,0)%
\includegraphics{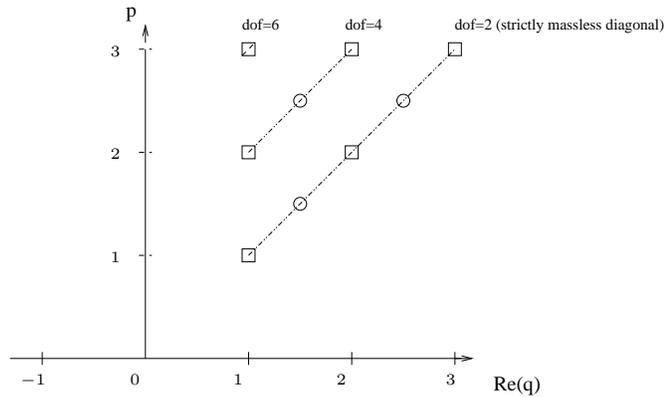}%
\end{picture}%
\setlength{\unitlength}{2131sp}%
\begingroup\makeatletter\ifx\SetFigFont\undefined
\def\x#1#2#3#4#5#6#7\relax{\def\x{#1#2#3#4#5#6}}%
\expandafter\x\fmtname xxxxxx\relax \def\y{splain}%
\ifx\x\y   
\gdef\SetFigFont#1#2#3{%
  \ifnum #1<17\tiny\else \ifnum #1<20\small\else
  \ifnum #1<24\normalsize\else \ifnum #1<29\large\else
  \ifnum #1<34\Large\else \ifnum #1<41\LARGE\else
     \huge\fi\fi\fi\fi\fi\fi
  \csname #3\endcsname}%
\else \gdef\SetFigFont#1#2#3{\begingroup
  \count@#1\relax \ifnum 25<\count@\count@25\fi
  \def\x{\endgroup\@setsize\SetFigFont{#2pt}}%
  \expandafter\x
    \csname \romannumeral\the\count@ pt\expandafter\endcsname
    \csname @\romannumeral\the\count@ pt\endcsname
  \csname #3\endcsname}%
\fi \fi\endgroup
\begin{picture}(5637,4561)(214,-4403)
\put(1426,-2836){\makebox(0,0)[lb]{\smash{\SetFigFont{6}{7.2}{rm}$1$}}}
\put(1426,-1636){\makebox(0,0)[lb]{\smash{\SetFigFont{6}{7.2}{rm}$2$}}}
\put(1426,-436){\makebox(0,0)[lb]{\smash{\SetFigFont{6}{7.2}{rm}$3$}}}
\put(376,-4261){\makebox(0,0)[lb]{\smash{\SetFigFont{6}{7.2}{rm}$-1$}}}
\put(1651,-4261){\makebox(0,0)[lb]{\smash{\SetFigFont{6}{7.2}{rm}$0$}}}
\put(2851,-4261){\makebox(0,0)[lb]{\smash{\SetFigFont{6}{7.2}{rm}$1$}}}
\put(4051,-4261){\makebox(0,0)[lb]{\smash{\SetFigFont{6}{7.2}{rm}$2$}}}
\put(5326,-4261){\makebox(0,0)[lb]{\smash{\SetFigFont{6}{7.2}{rm}$3$}}}
\end{picture}

\caption{Partially and strictly massless fields for integer and semi integer
spins up to spin $3$.}\label{f8}
\end{center}
\end{figure}

 It appears simple and  consistent to us to always
use definition $(\ref{mass1})$ which leads to partially massless
fields in dS space for integer as well as semi-integer spins.
These are represented together on Figure \ref{f8}. We can characterize
the partially   and  strictly massless lines by remarking that
they all correspond to members of the discrete series of
representation (except for the value $q=1/2$ which is contiguous
to the principal series). Thus there seems to be something very
special about the discrete series in relation with gauge
invariance.

\begin{figure}[h]
\begin{center}

\begin{picture}(0,0)%
\includegraphics{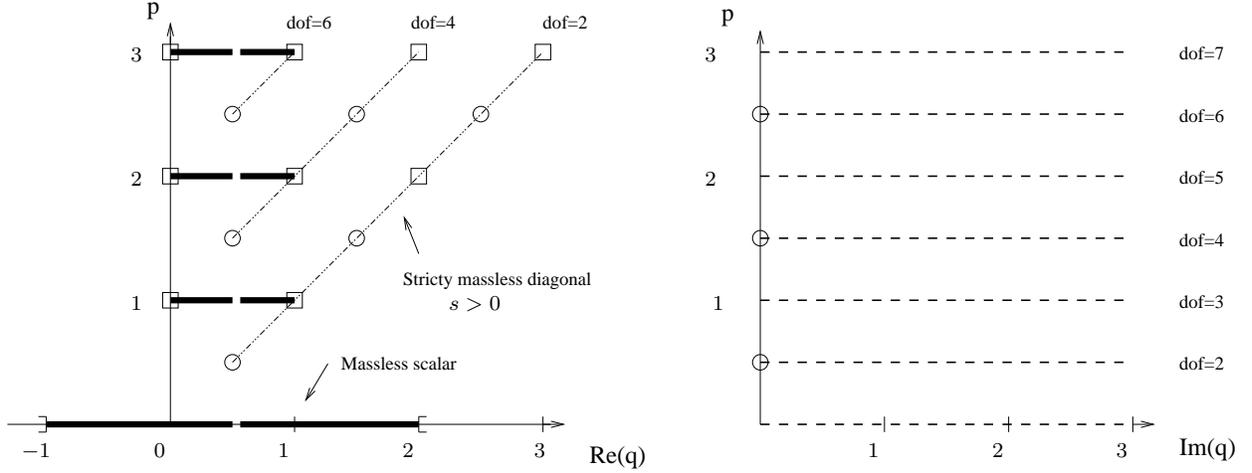}%
\end{picture}%
\setlength{\unitlength}{2565sp}%
\begingroup\makeatletter\ifx\SetFigFont\undefined
\def\x#1#2#3#4#5#6#7\relax{\def\x{#1#2#3#4#5#6}}%
\expandafter\x\fmtname xxxxxx\relax \def\y{splain}%
\ifx\x\y   
\gdef\SetFigFont#1#2#3{%
  \ifnum #1<17\tiny\else \ifnum #1<20\small\else
  \ifnum #1<24\normalsize\else \ifnum #1<29\large\else
  \ifnum #1<34\Large\else \ifnum #1<41\LARGE\else
     \huge\fi\fi\fi\fi\fi\fi
  \csname #3\endcsname}%
\else \gdef\SetFigFont#1#2#3{\begingroup
  \count@#1\relax \ifnum 25<\count@\count@25\fi
  \def\x{\endgroup\@setsize\SetFigFont{#2pt}}%
  \expandafter\x
    \csname \romannumeral\the\count@ pt\expandafter\endcsname
    \csname @\romannumeral\the\count@ pt\endcsname
  \csname #3\endcsname}%
\fi \fi\endgroup
\begin{picture}(11337,4561)(214,-4403)
\put(4501,-2836){\makebox(0,0)[lb]{\smash{\SetFigFont{8}{9.6}{rm}$s>0$}}}
\put(1426,-2836){\makebox(0,0)[lb]{\smash{\SetFigFont{8}{9.6}{rm}$1$}}}
\put(1426,-1636){\makebox(0,0)[lb]{\smash{\SetFigFont{8}{9.6}{rm}$2$}}}
\put(1426,-436){\makebox(0,0)[lb]{\smash{\SetFigFont{8}{9.6}{rm}$3$}}}
\put(8551,-4261){\makebox(0,0)[lb]{\smash{\SetFigFont{8}{9.6}{rm}$1$}}}
\put(9751,-4261){\makebox(0,0)[lb]{\smash{\SetFigFont{8}{9.6}{rm}$2$}}}
\put(10951,-4261){\makebox(0,0)[lb]{\smash{\SetFigFont{8}{9.6}{rm}$3$}}}
\put(7051,-2836){\makebox(0,0)[lb]{\smash{\SetFigFont{8}{9.6}{rm}$1$}}}
\put(6976,-1636){\makebox(0,0)[lb]{\smash{\SetFigFont{8}{9.6}{rm}$2$}}}
\put(6976,-436){\makebox(0,0)[lb]{\smash{\SetFigFont{8}{9.6}{rm}$3$}}}
\put(376,-4261){\makebox(0,0)[lb]{\smash{\SetFigFont{8}{9.6}{rm}$-1$}}}
\put(1651,-4261){\makebox(0,0)[lb]{\smash{\SetFigFont{8}{9.6}{rm}$0$}}}
\put(2851,-4261){\makebox(0,0)[lb]{\smash{\SetFigFont{8}{9.6}{rm}$1$}}}
\put(4051,-4261){\makebox(0,0)[lb]{\smash{\SetFigFont{8}{9.6}{rm}$2$}}}
\put(5326,-4261){\makebox(0,0)[lb]{\smash{\SetFigFont{8}{9.6}{rm}$3$}}}
\end{picture}

\caption{De Sitter UIR diagram up to p$=3$.}\label{f9}
\end{center}
\end{figure}

Although this comparison has only  been done for spins up to $3$,
it is clear that the UIR's figure will behave in the same way for
$p>3$ and one can expect that also in these cases the members of
discrete series of representations will correspond to partially
massless lines. Since  the mass relation (\ref{mass1}) can be
written
\begin{equation}\label{mass5}
m_{H}^{2}= \left[p(p-1)-q(q-1)\right]H^2=
\left[(p-q)(p+q-1)\right]H^2 \qquad\mbox{for $p>0$}\,,
\end{equation}
one could conjecture that  the partially massless fields will
assume the values given by (\ref{mass5}) with $q=p-1,..,1$ or
$3/2$. Of course, this still must be worked out properly. In
Figure \ref{f9}, we give a complete picture of the various fields up
to spin $3$ with the corresponding degrees of freedom.

Let us at last compute the mass values for the borderlines of the
unitary regions in order to view how these regions evolve with
increasing $p$. The ultimate  discrete series representation
values before the unitary regions (which starts with the
complementary series in the integer spin cases and directly with
the principal series in the half integer cases) correspond to
$q=1$ and $q=1/2$ for the integer and half integer cases
respectively. Thus, according to (\ref{mass5}) and using
$\Lambda=3H^{2}$ one gets
$$ \Lambda=\frac{3m^{2}_{H}}{p(p-1)}\quad\mbox{for the integer case,}
\qquad\Lambda=\frac{3m^{2}_{H}}{(p-\frac{1}{2})^2}\quad\mbox{for
the half integer case.}$$ It is clear that the allowed unitary
region for both cases approaches the region $\Lambda=0$ for
increasing $p$ (spin!). This phenomenon has already  been pointed
out in \cite{deser4}  with the difference that the fermion higher
spin field unitary region approaches the region around $\Lambda=0$
from below $\Lambda<0$, since these fields correspond to AdS
fields. Apart from this, we agree with S. Deser and A. Waldron on
the fact that as higher spin fields are included, the overlap
between the allowed unitary region narrows to the value of
$\Lambda=0$. This ``provides a sample of how particle kinematics
can be affected  by cosmological backgrounds allowing only certain
partial gauge theories and a restricted range of $m^{2}_{H}$ for a
given $\Lambda$".

\subsection{The mass of the graviton}
An interesting approach to the spin-$2$ field theory is given in
\cite{novello2} where the use of the so-called Fierz
representation allows to deal with the consistency problems
inherent to spin-$2$ coupling which originate in  the non
commutative nature of the covariant derivatives. The ambiguities
arise naturally since ones deals with second order derivatives
quantities. The difficulties encountered for general spin-$2$
coupling (to gravity or other fields) consist in  finding enough
conditions (analogous to the divergencelessness and tracelessness
condition in the flat case) in order  to reduce the degrees of
freedom of the field to pure spin-$2$ (for a clear review see
\cite{hindawi}). These conditions which must be consistent with
the equation of motion can be derived at least in the massive
case. In the massless case, consistency is related to the presence
of a local gauge invariance.

Applying the Fierz representation approach to the massive spin-$2$
field in (A)dS space, M. Novello and R. P. Neves show that the
graviton (helicity $\pm 2$) mass in AdS space is related to the cosmological constant
by  $m_{\sss NN}^{2}=-2\Lambda/3$ where $m_{\sss NN}$ is the mass used in \cite{novello1}. The origin of this non zero value of
the mass parameter for the graviton is related to the new form of
the field equation given in the Fierz formalism. This equation is
given by
\begin{equation}\label{none}
\hat{G}_{\mu\nu}+\frac{1}{2}m_{\sss NN}^{2}\left(  h_{\mu\nu}-g_{\mu\nu}
h\right)=0\,,\qquad\mbox{with}\qquad\hat{G}_{\mu\nu}=\frac{1}{2}\left({G}^{(a)}_{\mu\nu}+{G}^{(b)}_{\mu\nu}\right)\,,
\end{equation}
where
\begin{eqnarray}
{G}^{(a)}_{\mu\nu}=\frac{1}{2}\left[ \Box_{H}
h_{\mu\nu}+\nabla_{\mu}\nabla_{\nu}h-\left( \nabla_{\mu}
\nabla^{\lambda} h_{\nu\lambda}+\nabla_{\nu} \nabla^{\lambda}
h_{\mu\lambda}\right)-g_{\mu\nu}\left( \Box_{H} h -
\nabla_{\lambda}\nabla_{\rho}h^{\lambda\rho}\right)\right]\,,
\nonumber\\
{G}^{(b)}_{\mu\nu}=\frac{1}{2}\left[ \Box_{H}
h_{\mu\nu}+\nabla_{\mu}\nabla_{\nu}h-\left( \nabla^{\lambda}
\nabla_{\mu} h_{\nu\lambda}+\nabla^{\lambda} \nabla_{\nu}
h_{\mu\lambda}\right)-g_{\mu\nu}\left( \Box_{H} h -
\nabla_{\lambda}\nabla_{\rho}h^{\lambda\rho}\right)\right]\,,
\end{eqnarray}
which is symmetric with respect to the ordering of the covariant
derivatives.  Now since
\begin{equation}
\nabla_{\rho}\nabla_{\lambda}h_{\mu\nu}-
\nabla_{\lambda}\nabla_{\rho} h_{\mu\nu}=-H^{2}\left(
g_{\rho\mu}h_{\lambda\nu}+g_{\rho\nu}
h_{\lambda\mu}-g_{\lambda\mu}h_{\rho\nu}-g_{\lambda\nu}
h_{\rho\mu}\right)\,,
\end{equation}
which yields
\begin{equation}
\nabla^{\rho}\nabla_{\mu}h_{\rho\nu}= \nabla_{\mu}\nabla^ {\rho}
h_{\rho\nu}-4H^{2} h_{\mu\nu}+H^{2}g_{\mu\nu}h\,,
\end{equation}
one finally gets the field equation
\begin{eqnarray}
(\Box_H&+&4H^2)h_{\mu\nu}-g_{\mu\nu}(\Box_H+H^2) h-\left(
\nabla_{\mu} \nabla\cdot h_{\nu}+\nabla_{\nu} \nabla\cdot
h_{\mu}\right) \nonumber\\
&+&g_{\mu\nu}\nabla^{\lambda}\nabla^{\rho}h_{\lambda\rho}
+\nabla_{\mu}\nabla_{\nu}h+m_{\sss NN}^{2}\left(h_{\mu\nu }-h
g_{\mu\nu}\right)=0\,.
\end{eqnarray}
The traditional gauge invariant field equation of the graviton
(based on ${G}^{(a)}_{\mu\nu}$ ) is given by
\begin{eqnarray}\label{graviton}
(\Box_H+2H^2)h_{\mu\nu}-g_{\mu\nu}(\Box_H-H^2) h-\left(
\nabla_{\mu} \nabla\cdot h_{\nu}+\nabla_{\nu} \nabla\cdot
h_{\mu}\right)+g_{\mu\nu}\nabla^{\lambda}\nabla^{\rho}h_{\lambda\rho}
+\nabla_{\mu}\nabla_{\nu}h=0\,,\nonumber
\end{eqnarray}
and is equivalent to the former when $m_{\sss NN}^{2}=-2\Lambda/3$. This
equation can  be transcribed  to ambient space notations using the
relations (\ref{passage1}), (\ref{passage2}) and (\ref{pass}). One  finds that
it reads
\begin{equation}\label{sp}
\left(Q^{(1)}+6\right){\cal K}(x)+D_2\partial_2 \cdot {\cal K}(x)=0\,,
\end{equation}
where the operators $D_2$ et $\partial_2 $ are respectively the
generalized gradient and divergence
\begin{equation} D_2{\cal K}=H^{-2}{\cal S}(\bar
\partial-H^2x){\cal K},\qquad\partial_2\cdot
{\cal K}=\partial \cdot{\cal K}- H^2 x {\cal K}'-\frac{1}{2} \bar
\partial {\cal K}'\,.\end{equation}
with ${\cal S}$ a symmetrization operator.
If we compare  Eq. $(\ref{sp})$ with the eigenvalue equation
$$\left( Q^{(1)}-\langle Q^{(1)}\rangle\right){\cal K}=0\,,$$
it is
found that the relevant physical subspace (which can be associated
to an UIR) is made of the solution of
\begin{equation}
\left(Q^{(1)}+6\right){\cal K}(x)=0\,,
\end{equation}
and thus corresponds to the discrete series of unitary
representations with $p=q=2$ ($\langle Q^{(1)}\rangle=-6$). Thus we see that the field associated to the mass $m^{2}_{\sss NN}=-2\Lambda/3$ and  verifying Eq. (\ref{none}) corresponds to the usual strictly massless UIR of the dS group. Precisely the one which corresponds to $m_H^2=0$ and where the gauge invariance allows to reduce the degrees of freedom to two (helicity $\pm 2$).

Of course the problem is that in dS space, the value of $m_{\sss NN}$ would be imaginary because $\Lambda=3H^2$. Let us compare $m_{NN}$ with our mass definition $m_H^2$ in order to understand this problem. It is clear (since $m_H^2=0$ when $m_{\sss NN}^2$) that $m_{\sss NN}^{2}$ is related to $m_{H}^{2}$  through the formula
\begin{equation}
m_{H}^{2}=m_{\sss NN}^{2}+\frac{2\Lambda}{3}\,.
\end{equation}
Equivalently using $(\ref{mass1})$ one has
\begin{equation}
m_{\sss NN}^{2}=m_{H}^{2}-\frac{2\Lambda}{3}=H^{2}\langle
Q_{}^{(1)}\rangle+4H^{2}.
\end{equation}
It is obvious from that relation that $m_{\sss NN}^2\in\setR^+$
fails to describe all the UIR's of the de Sitter group. In
particular,  {\bf the UIR with $p=q=2$ and $\langle
Q_{}^{(1)}\rangle=-6 $ which is the strictly massless UIR is
eliminated}. We would like to insist on the fact that we agree
with the authors when they claim that the field equation
(\ref{graviton}) is the graviton field equation. But whereas in
\cite{novello1} the corresponding field is not a dS field (because
of the imaginary mass) on the contrary our mass definition entails
that it is a dS field, which seems reasonable since after all
$\langle Q_{p=q=2}^{(1)}\rangle$ is a dS unitary representation
(not AdS).

\section{Outlooks}
This mass definition $m_H^2$ and the ($p,q$) diagrams enabled us
to locate the fields where gauge invariance appears  as members of
the discrete series of representation. This has been done up to
p=3. Although it must still be rigourously shown, it is reasonable
to believe that also for values of $p>3$, these representations
will correspond to fields featuring gauge invariance. The
challenge therefore is to establish the precise relation between
the discrete series of UIR's and the property of gauge invariance.

\section*{ACKNOWLEDGMENTS}
The author thanks J. Renaud, J-P. Gazeau and E. Huguet  for
valuables discussions and helpful criticisms.
\begin{appendix}
\section{D'Alembertian operator in ambient space notations}
In the following, we would like to prove that given   a symmetric
transverse tensor $h_{\lambda_{1}..\lambda_{r}}(x)$  of rank $r$,
and linked to the ambient space tensor ${\cal
K}_{\beta_{1}..\beta_{r}}(x)$ by
\begin{equation}\label{passage11}
 h_{\lambda_{1}..\lambda_{r}}(X)=\frac{\partial
x^{\beta_{1}}}{\partial X^{\lambda_{1}}} ...\frac{\partial
x^{\beta_{r}}} {\partial X^{\lambda_{r}}}{\cal
K}_{\beta_{1}..\beta _{r}}(x)\,,
\end{equation}
the action of the d'Alembertian in local coordinates is given by
\begin{eqnarray}\label{passa}
\Box_{H}h_{\lambda_{1}..\lambda_{r}}(X)&=&\frac{\partial
x^{\beta_{1}}}{\partial X^{\lambda_{1}}}.. \frac{\partial
x^{\beta_{r}}} {\partial X^{\lambda_{r}}}\Big{[}-H^{2}\left(
Q^{(1)}_{0}+r\right){\cal K}_{\beta_{1}..\beta_{r}}
+2H^{4}\sum_{j=1}^{r}x_{\beta_{j}}\sum_{i<j}^{}x_{\beta_{i}} {\cal
K
}_{\beta_{1}..\hat{\beta_{i}}\hat{\beta_{j}}..\beta_{r}}'\nonumber\\
&-&2H^{2}\sum_{i=1}^{r}x_{\beta_{i}}\left( \bar\partial\cdot{\cal
K }_{\beta_{1}..\hat{\beta_{i}}..\beta_{r}}-H^{2}x\cdot{\cal K
}_{\beta_{1}..\hat{\beta_{i}}..\beta_{r}} \right)\Big{]}\,.
\end{eqnarray}
For this, we must  compute the expression
\begin{equation}\label{passage22}
\nabla_{\mu}\nabla_{\nu}h_{\lambda_{1}..\lambda_{r}}=
\frac{\partial x^\alpha}{\partial X^\mu} \frac{\partial x^\gamma}
{\partial X^\nu} \frac{\partial x^{\beta_{1}}}{\partial
X^{\lambda_{1}}} ...\frac{\partial x^{\beta_{r}}} {\partial
X^{\lambda_{r}}}
\mbox{Trpr}\bar{\partial}_{\alpha}\mbox{Trpr}\bar{\partial}_{\gamma}
{\cal K}_{\beta_{1}..\beta_{r}}\,,
\end{equation}
where the transverse projection operator is defined by
\begin{equation} \left(\mbox{Trpr}
\,{\cal
K}\right)_{\lambda_{1}..\lambda_{r}}\equiv\theta^{\beta_{1}}_{\lambda_{1}}
..\theta^{\beta_{r}}_{\lambda_{r}}{\cal
 K}_{\beta_{1}..\beta_{r}}\,.\nonumber
\end{equation}
The first step is to prove that
\begin{eqnarray}\label{tr1}
\left(\mbox{Trpr}\bar{\partial}{\cal
K}\right)_{\alpha\beta_{1}..\beta_{l}}&=&\bar\partial_{\alpha}{\cal
K }_{\beta_{1}..\beta_{r}}-H^{2}\sum_{i=1}^{r}x_{\beta_{i}}{\cal K
}_{\beta_{1}..\hat{\beta_{i}}..\beta_{r}\alpha}\,,\\
\left(\mbox{Trpr }\bar{\partial}\mbox{Trpr}\bar{\partial}{\cal K
}\right)_{\alpha\gamma\beta_{1}..\beta_{r}}&=&
\bar\partial_{\alpha}\bar\partial_{\gamma}{\cal K
}_{\beta_{1}..\beta_{r}}-H^{2}\sum_{i=1}^{r}\theta_{\alpha\beta_{i}}{\cal
K}_{\beta_{1}..\hat{\beta_{i}}..\beta_{r}\gamma}+2H^{4}\sum_{j=1}^{r}x_{\beta_{j}}\sum_{i<j}^{}x_{\beta_{i}}
{\cal K
}_{\beta_{1}..\hat{\beta_{i}}\hat{\beta_{j}}..\beta_{l}\alpha\gamma}\nonumber\\
&-&H^{2}x_{\gamma}\bar\partial_{\alpha}{\cal
K}_{\beta_{1}..\beta_{r}} -H^{2}\sum_{i=1}^{r}x_{\beta_{i}}{\cal
S}_{\alpha\gamma}\left( \bar\partial_{\alpha}{\cal K
}_{\beta_{1}..\hat{\beta_{i}}..\beta_{r}\gamma}-H^{2}x_{\alpha}{\cal
K }_{\beta_{1}..\hat{\beta_{i}}..\beta_{r}\gamma}
\right)\label{tr2} \,.
\end{eqnarray}
eIt is  straightforward using
$\theta_{\beta_{1}}^{\beta_{1}'}..\,\theta_{\beta_{r}}^{\beta_{r}'}
{\cal K }_{\beta_{1}'..\beta_{r}'}={\cal K
}_{\beta_{1}..\beta_{r}}$ and  expressions such as $x\cdot
\bar\partial=0,\;x\cdot\theta=0,\;x\cdot{\cal K}=0,\,\bar\partial
x=\theta,\; $ $\bar\partial_{\alpha
}\theta_{\beta\gamma}=H^{2}\theta_{\alpha\beta}x_{\gamma}+H^{2}\theta_{\alpha\gamma}x_{\beta}$
to establish the relation  (\ref{tr1}) :
\begin{eqnarray}
\left(\mbox{Trpr}\bar{\partial}{\cal
K}\right)_{\alpha\beta_{1}..\beta_{r}}&=&\theta_{\beta_{1}}^{\beta_{1}'}..\theta_{\beta_{r}}^{\beta_{r}'}
\bar\partial_{\alpha}{\cal K
}_{\beta_{1}'..\beta_{r}'}=\theta_{\beta_{1}}^{\beta_{1}'}..\theta_{\beta_{r-1}}^{\beta_{r-1}'}\left(
\bar\partial_{\alpha}{\cal K }_{\beta_{1}'..\beta_{r-1}'\beta_{r}}
-H^{2}x_{\beta_{r}}{\cal K
}_{\beta_{1}'..\beta_{r-1}'\alpha}\right) \nonumber\\
&=&\theta_{\beta_{1}}^{\beta_{1}'}..\theta_{\beta_{r-1}}^{\beta_{r-1}'}
\bar\partial_{\alpha}{\cal K }_{\beta_{1}'..\beta_{r-1}'\beta_{r}}
-H^{2}x_{\beta_{r}}{\cal K
}_{\beta_{1}..\beta_{r-1}\alpha}\nonumber \\
&=&\bar\partial_{\alpha}{\cal K
}_{\beta_{1}..\beta_{r}}-H^{2}\sum_{i=1}^{r}x_{\beta_{i}}{\cal K
}_{\beta_{1}..\hat{\beta_{i}}..\beta_{r}\alpha}\,.
\end{eqnarray}
Let us now prove formula (\ref{tr2}). First of all, given
(\ref{tr1}) one has
\begin{eqnarray}
\left(\mbox{Trpr }\bar{\partial}\mbox{Trpr}\bar{\partial}{\cal K
}\right)_{\alpha\gamma\beta_{1}..\beta_{r}} = \mbox{Trpr
}\bar{\partial}\left( \bar\partial {\cal K}-H^2\sum x {\cal
K}\right)=
 \left(\mbox{Trpr
}\bar{\partial}\bar{\partial}{\cal K
}\right)_{\alpha\gamma\beta_{1}..\beta_{r}}-H^{2} \left(\mbox{Trpr
}\bar{\partial}\,\sum x{\cal K
}\right)_{\alpha\gamma\beta_{1}..\beta_{r}} \nonumber\,,
\end{eqnarray}
and since $\theta\cdot x=0$ one gets
\begin{eqnarray}
\left(\mbox{Trpr }\bar{\partial}\mbox{Trpr}\bar{\partial}{\cal K
}\right)_{\alpha\gamma\beta_{1}..\beta_{r}}
=\theta_{\gamma}^{\gamma'}\theta_{\beta_{1}}^{\beta_{1}'}..\theta_{\beta_{r}}^{\beta_{r}'}
\bar\partial_{\alpha}\bar\partial_{\gamma'}{\cal K
}_{\beta_{1}'..\beta_{r}'}
-H^{2}\sum_{i=1}^{r}\theta_{\alpha\beta_{i}}{\cal K
}_{\beta_{1}..\hat{\beta_{i}}..\beta_{r}\gamma} \,.
\end{eqnarray}
Moreover one has
\begin{eqnarray}
\theta_{\gamma}^{\gamma'}\theta_{\beta_{1}}^{\beta_{1}'}..\theta_{\beta_{r}}^{\beta_{r}'}
\bar\partial_{\alpha}\bar\partial_{\gamma'}{\cal K
}_{\beta_{1}'..\beta_{r}'}
&=&\theta_{\beta_{1}}^{\beta_{1}'}..\theta_{\beta_{r}}^{\beta_{r}'}\left(
\bar\partial_{\alpha}\bar\partial_{\gamma}{\cal K
}_{\beta_{1}'..\beta_{r}'}-H^{2}x_{\gamma}\bar\partial_{\alpha}{\cal
K }_{\beta_{1}'..\beta_{r}'}
\right)\nonumber\\
&=&\theta_{\beta_{1}}^{\beta_{1}'}..\theta_{\beta_{r}}^{\beta_{r}'}
\bar\partial_{\alpha}\bar\partial_{\gamma}{\cal K
}_{\beta_{1}'..\beta_{r}'}-H^{2}x_{\gamma}\left(\theta_{\beta_{1}}^{\beta_{1}'}..\theta_{\beta_{r}}^{\beta_{r}'}
\bar\partial_{\alpha}{\cal K }_{\beta_{1}'..\beta_{r}'} \right)\nonumber\\
&=&\theta_{\beta_{1}}^{\beta_{1}'}..\theta_{\beta_{r}}^{\beta_{r}'}
\bar\partial_{\alpha}\bar\partial_{\gamma}{\cal K
}_{\beta_{1}'..\beta_{r}'}-H^{2}x_{\gamma}\left(\bar\partial_{\alpha}
{\cal K
}_{\beta_{1}..\beta_{r}}-H^{2}\sum_{i}^{r}x_{\beta_{i}}{\cal K
}_{\beta_{1}..\hat{\beta_{i}}..\beta_{r}\alpha}  \right)\,.
\end{eqnarray}
Finally we must calculate
$\theta_{\beta_{1}}^{\beta_{1}'}..\theta_{\beta_{r}}^{\beta_{r}'}
\bar\partial_{\alpha}\bar\partial_{\gamma}{\cal K
}_{\beta_{1}'..\beta_{r}'}$. The procedure here is to develop the
derivatives of $\theta_{\beta_r}^{\beta_{r}'}$ in expressions like
$$\bar\partial_{\alpha}\bar\partial_{\gamma}{\cal K
}_{\beta_{1}'..\beta_{r}}=\bar\partial_{\alpha}\bar\partial_{\gamma}\,\theta_{\beta_r}^{\beta_{r}'}\,{\cal
K }_{\beta_{1}'..\beta_{r}'}\,.$$ In this way, after a tedious but
simple computation one gets
\begin{eqnarray}
\theta_{\beta_{r}}^{\beta_{r}'}
\bar\partial_{\alpha}\bar\partial_{\gamma}{\cal K
}_{\beta_{1}'..\beta_{r}'} &=&
\bar\partial_{\alpha}\bar\partial_{\gamma}{\cal K
}_{\beta_{1}'..\beta_{r-1}'\beta_{r}}+H^{4}x_{\alpha}x_{\beta_{r}}{\cal
K }_{\beta_{1}'..\beta_{r-1}'\gamma}-H^{2}x_{\beta_{r}}{\cal S
}_{\alpha\gamma}\bar\partial_{\alpha}{\cal K
}_{\beta_{1}'..\beta_{r-1}'\gamma}\nonumber\,,
\end{eqnarray}
where ${\cal S }_{\alpha\gamma}$ is the non normalized
symmetrization operator with respect to $\alpha$ and $\gamma $.
Thus finally one obtains
\begin{eqnarray}
\theta_{\beta_{1}}^{\beta_{1}'}..\theta_{\beta_{r}}^{\beta_{r}'}
\bar\partial_{\alpha}\bar\partial_{\gamma}{\cal K
}_{\beta_{1}'..\beta_{r}'}
&=&\theta_{\beta_{1}}^{\beta_{1}'}..\theta_{\beta_{r-1}}^{\beta_{r-1}'}\left(
\bar\partial_{\alpha}\bar\partial_{\gamma}{\cal K
}_{\beta_{1}'..\beta_{r-1}'\beta_{r}}+H^{4}x_{\alpha}x_{\beta_{r}}{\cal
K }_{\beta_{1}'..\beta_{r-1}'\gamma}-H^{2}x_{\beta_{r}}{\cal S
}_{\alpha\gamma}\bar\partial_{\alpha}{\cal K
}_{\beta_{1}'..\beta_{r-1}'\gamma}
\right)\nonumber\\
&=&\theta_{\beta_{1}}^{\beta_{1}'}..\theta_{\beta_{r-1}}^{\beta_{r-1}'}
\bar\partial_{\alpha}\bar\partial_{\gamma}{\cal K
}_{\beta_{1}'..\beta_{r-1}'\beta_{r}}+H^{4}x_{\alpha}x_{\beta_{r}}{\cal
K }_{\beta_{1}..\beta_{r-1}\gamma}  \nonumber\\
&-&H^{2}x_{\beta_{r}}{\cal S
}_{\alpha\gamma}\left(\bar\partial_{\alpha}{\cal K
}_{\beta_{1}..\beta_{r-1}\gamma}-H^{2}\sum_{j=1}^{r-1}x_{\beta_{j}}{\cal
K}_{\beta_{1}..\hat{\beta_{j}}..\beta_{r-1}\alpha\gamma}\right)
 \nonumber\\
&=&\bar\partial_{\alpha}\bar\partial_{\gamma}{\cal K
}_{\beta_{1}..\beta_{r}}+H^{4}x_{\alpha}\sum_{j=1}^{r}x_{\beta_{j}}
{\cal K }_{\beta_{1}..\hat{\beta_{j}}..\beta_r\gamma}
\nonumber\\&-&H^{2}\sum_{j=1}^{r}x_{\beta_{j}}{\cal S
}_{\alpha\gamma}\left(\bar\partial_{\alpha}{\cal K
}_{\beta_{1}..\hat{\beta_{j}}..\beta_{r}\gamma}-
H^{2}\sum_{i<j}^{}x_{\beta_{i}}{\cal K
}_{\beta_{1}..\hat{\beta_{i}}\hat{\beta_{j}}..\alpha\gamma}\,
\right)\,,
\end{eqnarray}
which completes the proof of formula (\ref{tr2}). We are now in
position to compute the d'Alembertian operator on any transverse
tensor. Note that it is easy to show that the  metric in local
coordinates $g_{\mu\nu}^{}$ corresponds through
$(\ref{passage11})$ to  the transverse projector $\theta=\eta+H^2
x x $ and therefore we have
\begin{equation}\label{}
\Box_{H} h_{\lambda_{1}..\lambda_{r}}= \frac{\partial
x^{\beta_{1}}}{\partial X^{\lambda_{1}}} ...\frac{\partial
x^{\beta_{r}}} {\partial
X^{\lambda_{r}}}\;\theta^{\alpha\gamma}\;\left(
\mbox{Trpr}\bar{\partial}_{}\mbox{Trpr}\bar{\partial}_{} {\cal
K}_{}\right)_{\alpha\gamma\beta_{1}..\beta_{r}}\,.
\end{equation}
Formula $(\ref{passa})$ then follows from $(\ref{tr2})$.
\end{appendix}


\begin{thebibliography}{99}


\bibitem{eric} T. Garidi, E. Huguet, J. Renaud,  Phys. Rev. D {\bf 67}, 124028 (2003).

\bibitem{birrel}
Birrel N. D., Davies P. C. W., Cambridge Monoghraphs on
Mathematical Physics, $(1984)$, {\it Quantum fields in curved
space}.

\bibitem{deser1} S. Deser, R.I. Nepomechie, Phys. Lett. B, {\bf 132}, 321 (1983).

\bibitem{deser2} S. Deser, R.I. Nepomechie, Annals Phys. {\bf 154}, 396 (1984).

\bibitem{higu2}  A. Higuchi,  Nucl. Phys. B {\bf 282},   397  (1987).

\bibitem{deser3} S. Deser, A. Waldron, Phys. Rev. Lett. {\bf 87},  031601 (2001).

\bibitem{deser4} S. Deser, A. Waldron, Nucl. Phys. B {\bf 607},  577-604  (2001).

\bibitem{deser5} S. Deser, A. Waldron, Phys. Lett. B {\bf 508}, 347-353 (2001).
\bibitem{deser6} S. Deser, A. Waldron, Phys. Lett. B {\bf 513}, 137-141  (2001).
\bibitem{novello1}   M. Novello,  R.P. Neves,
Class. Quant. Grav. {\bf 20},  L67- L73 (2003).
\bibitem{dewitt} B. S. De Witt,  {\it Relativity, Groups and Topology}
Gordon $\&$ Breach, New York (1964).
\bibitem{gareta1}  Gazeau J. P., Renaud J., Takook M. V.,
Class. Quant. Grav. $17 \,(2000) \, $L$1415- $L$1434$.
\bibitem{dix} J. Dixmier, Bull. Soc. Math. France, {\bf 89} 9 (1961).
\bibitem{tak}  B. Takahashi, Bull. Soc. Math. France {\bf 91}, 289
(1963).
\bibitem{fr} C. Fronsdal, Phys. Rev. D, {\bf 20} 848  (1979).
\bibitem{gaha} J. P. Gazeau,  M. Hans, J. Math. Phys.,
{\bf 29}  2533  (1988).
\bibitem{di} P. A. M., Dirac  Ann.  Math., {\bf 36}  657 (1935).
\bibitem{mini} J. Mickelsson, J. Niederle, Comm. Math. Phys. {\bf 27} 167 (1972).
\bibitem{lipsman} L. Lipsman, Springer Verlag, {\it Group
Representations}, Lecture Notes in Mathematics, 388 (1974).



\bibitem{barutbohm} A. O. Barut, A. B\"{o}hm ,  J. Math. Phys. {\bf 11} 2938 (1970).

\bibitem{anflafrons}  E. Angelopoulos,  M. Flato,  C. Fronsdal,
D. Sternheimer, Phys. Rev . D {\bf 23}, 1278 (1981).
\bibitem{gahamu} J. P. Gazeau,  M. Hans,  R. Murenzi,
Class. Quantum Grav. {\bf 6},  329 (1989).

\bibitem{lesimple} Lesimple M.,
Letters Math. Phys., $15\;(1988) \;143$.
\bibitem{gagata}  T. Garidi,   J.P. Gazeau, M.V. Takook,  J. Math.
Phys., {\bf 44, 9} 3838 (2003), hep-th/0302022.
\bibitem{hindawi}  A. Hindawi,  B. A. Ovrut,  D. Waldram,
Phys. Rev. D {\bf 53}, 5583 (1996).

\bibitem{allen1} Allen B. Jacobson T. ,
Commun. Math. Phys., $103\, (1986) \,669$.

\bibitem{anilto2}   I. Antoniadis,   J. Iliopoulos,
 T. N. Tomaras, Nuclear Phys. B {\bf 462},  437 (1996).
\bibitem{novello2}   M. Novello,  R.P. Neves,
Class. Quant. Grav. {\bf 19},  5335-5351 (2002).

\end{thebibliography}
\end{document}